\documentclass[12pt]{article}
\usepackage{graphicx,epsf}

\newcommand{\lp}[1]{\includegraphics[width=0.13\textwidth]{D#1.ps}}

\def\ep{\epsilon}
\def\pslash{\rlap{\hspace{0.02cm}/}{p}}
\def\vslash{\rlap{\hspace{0.011cm}/}{v}}
\def\Dslash{\rlap{\hspace{0.10cm}/}{D}}
\def\bm#1{\mbox{\boldmath$#1$\unboldmath}}

\begin{document}

\begin{titlepage}

\begin{flushright}
CLNS~00/1713\\
{\tt hep-ph/0012183}
\end{flushright}

\vspace{1.5cm}
\begin{center}
\Large\bf\boldmath
Two-Loop Renormalization of Heavy--Light Currents\\
At Order $1/m_Q$ in the Heavy-Quark Expansion
\unboldmath
\end{center}

\vspace{0.5cm}
\begin{center}
Thomas Becher, Matthias Neubert and Alexey A. Petrov\\[0.1cm]
{\sl Newman Laboratory of Nuclear Studies, Cornell University\\
Ithaca, NY 14853, USA}
\end{center}

\vspace{1.5cm}
\begin{abstract}
\vspace{0.2cm}\noindent
We present exact results, at next-to-leading order in 
renormalization-group improved perturbation theory, for the Wilson 
coefficients appearing at order $1/m_Q$ in the heavy-quark expansion of 
heavy--light current operators. To this end, we complete the calculation 
of the corresponding two-loop anomalous dimension matrix. Our results 
are important for determinations of $|V_{ub}|$ using exclusive and
inclusive semileptonic $B$ decays. They are also relevant to 
computations of the decay constant $f_B$ based on a heavy-quark 
expansion.
\end{abstract}

\vspace{1.0cm}

\vspace{3.0cm}
\noindent
December 2000

\end{titlepage}

\section{Introduction}

Hadronic matrix elements of flavor-changing currents are of paramount 
importance to the phenomenology of weak decays. They appear, e.g., in
the theoretical description of semileptonic $B$ decays. These processes
give direct access to the elements $|V_{cb}|$ and $|V_{ub}|$ of the
quark-mixing matrix, provided the hadronic matrix elements are known 
from some nonperturbative computation. For inclusive semileptonic 
decays, the matrix elements can be calculated using a short-distance 
expansion in the large $b$-quark mass. Improving the theoretical 
accuracy in the determinations of $|V_{ub}|$ is currently one of the
most important challenges in heavy-flavor theory.

Relevant to the determination of $|V_{ub}|$ are hadronic matrix elements 
of the weak current $\bar u\gamma^\alpha(1-\gamma_5)b$. In general, local 
current operators composed of a heavy quark $Q$ and a light  
quark $q$ exhibit an interesting behavior under renormalization at 
scales below the heavy-quark mass. Large logarithms of the type 
$\alpha_s\ln(m_Q/\mu)$ arise from the exchange of gluons that are 
hard with respect to the light quark but soft with respect to the heavy 
quark. Since such gluons see the heavy quark as a static color source, 
their effects can be investigated systematically in the framework of 
the heavy-quark effective theory (HQET), which provides a systematic
expansion of matrix elements in inverse powers and logarithms of $m_Q$
\cite{review}. In the HQET, the four-component heavy-quark field is 
replaced by a velocity-dependent two-component field $h_v$ satisfying 
$\vslash\,h_v=h_v$, where $v$ is the four-velocity of the hadron 
containing the heavy quark. Operators in the effective theory have a 
different evolution than in usual QCD. For instance, whereas the vector 
current $\bar q\,\gamma^\alpha Q$ is conserved in QCD (i.e., its 
anomalous dimension vanishes), the corresponding current 
$\bar q\,\gamma^\alpha h_v$ in the HQET has a nontrivial anomalous 
dimension, which governs its renormalization-group (RG) evolution for 
scales below $m_Q$. 

As an example, consider the heavy-quark expansion of the vector
current (neglecting the light-quark mass $m_q$) \cite{FaGr,MN94}
\begin{equation}\label{vector}
   \bar q\,\gamma^\alpha Q \cong
   \sum_{i=1,2} C_i(\mu)\,J_i
   + \frac{1}{2m_Q} \sum_{j=1}^{10} B_j(\mu)\,Q_j
   + O(1/m_Q^2) \,,
\end{equation}
where the symbol $\cong$ is used to indicate that this is a relation 
that holds after taking matrix elements on both sides. (In a
regularization scheme with anticommuting $\gamma_5$, the expansion of 
the axial-vector current is obtained by replacing $\bar q\to
-\bar q\gamma_5$ in the HQET operators.) $J_1=\bar q\,\gamma^\alpha h_v$ 
and $J_2=\bar q\,v^\alpha h_v$ are the effective operators entering at 
leading order in the expansion. The operators $Q_j$ appearing at 
next-to-leading order (NLO) form a basis of dimension-4 operators in the 
HQET that closes under renormalization. The standard choice of these 
operators is (using the notation $iD=i\partial+g_s A$ and 
$(iD)^\dagger=-i\overleftarrow{\partial}+g_s A$)
\begin{eqnarray}\label{vbasis}
   Q_1 &=& \bar q\,\gamma^\alpha i\Dslash h_v \,,\hspace{1.5cm}
   Q_4 = \bar q\,(iv\!\cdot\!D)^\dagger\gamma^\alpha h_v \,,
    \nonumber\\
   Q_2 &=& \bar q\,v^\alpha i\Dslash h_v \,, \hspace{1.52cm}
   Q_5 = \bar q\,(iv\!\cdot\!D)^\dagger v^\alpha h_v \,, \nonumber\\
   Q_3 &=& \bar q\,iD^\alpha h_v \,, \hspace{1.75cm}
   Q_6 = \bar q\,(i D^\alpha)^\dagger h_v \,,
\end{eqnarray}
and
\begin{eqnarray}
   Q_{7,8} &=& i\!\int\!\mbox{d}^4x\,\mbox{T}\{
    J_{1,2}(0),\bar h_v (iD)^2 h_v(x) \} \,, \nonumber\\
   Q_{9,10} &=& i\!\int\!\mbox{d}^4x\,\mbox{T}\{
    J_{1,2}(0),\frac{g_s}{2}\,\bar h_v\sigma_{\mu\nu} G^{\mu\nu} h_v(x)
    \} \,.
\end{eqnarray}
In addition to the local operators $Q_{1\dots 6}$, there appear four
bilocal operators $Q_{7\dots 10}$ built from time-ordered products of
the leading-order currents with an insertion of the ``kinetic operator'' 
$\bar h_v (iD)^2 h_v$ or the ``chromo-magnetic operator''
$\frac{g_s}{2}\,\bar h_v\sigma_{\mu\nu} G^{\mu\nu} h_v$. These operators
appear at order $1/m_Q$ in the effective Lagrangian of the HQET. Part of
their effect is to take into account the expansion of the external 
hadron states in terms of states of the effective theory \cite{review}.
The Wilson coefficients of these nonlocal operators are given by the 
products of the coefficients of their component operators, i.e.,
\begin{equation}\label{B7to10}
   B_{7,8}(\mu) = C_{1,2}(\mu) \,, \qquad
   B_{9,10}(\mu) = C_{1,2}(\mu)\,C_{\rm mag}(\mu) \,,
\end{equation}
where $C_{\rm mag}(\mu)$ is the coefficient of the chromo-magnetic 
operator in the HQET Lagrangian, which has been calculated at NLO in 
\cite{ABN}. The kinetic operator is not renormalized \cite{LuMa}.

The calculation of the Wilson coefficients $C_i(\mu)$ and $B_j(\mu)$ at
NLO in RG-improved perturbation theory requires one-loop matching at the 
scale $\mu=m_Q$ and two-loop anomalous dimensions in the effective 
theory. The leading-order coefficients $C_i(\mu)$ have been computed 
long ago \cite{JiMu,BrGr}. The calculation at order $1/m_Q$ is, however, 
much more complicated. The leading logarithmic corrections were 
obtained in \cite{FaGr}. First steps towards a NLO analysis were made in 
\cite{MN94}, where the one-loop matching calculations were computed, and 
general arguments based on the reparameterization invariance \cite{LuMa} 
of the HQET were used to derive the exact relations (valid to all orders 
in perturbation theory)
\begin{equation}\label{C1C2}
   B_1(\mu) = C_1(\mu) \,, \qquad
   B_2(\mu) = \frac{1}{2}\,B_3(\mu) = C_2(\mu) \,.
\end{equation}
Moreover, it was shown that the $10\times 10$ anomalous dimension matrix
governing the mixing of the operators $Q_j$ under RG transformations has
the texture
\begin{equation}\label{gammamatr}
   \bm{\gamma}
   = \left( \begin{array}{ccc}
   \bm{\gamma_{\rm hl}}~ & \bm{\gamma_A} & 0 \\
   0 & \bm{\gamma_B} & 0 \\
   0 & \bm{\gamma_C} & ~\bm{\gamma_D}  
   \end{array} \right) ,
\end{equation}
where
\begin{eqnarray}\label{entries}
   \bm{\gamma_{\rm hl}}
   &=& \mbox{diag}(\gamma_1,\gamma_1,\gamma_1) \,, \nonumber\\
   \bm{\gamma_D}
   &=& \mbox{diag}(\gamma_1,\gamma_1,\gamma_1+\gamma_{\rm mag},
                   \gamma_1+\gamma_{\rm mag}) \,, \nonumber\\
   (\bm{\gamma_B})_{ij}
   &=& \gamma_1\,\delta_{ij} + \delta_{i3}\,(\bm{\gamma_A})_{3j} \,.
\end{eqnarray}
Here $\gamma_1$ and $\gamma_{\rm mag}$ are the anomalous dimensions of 
the leading-order heavy--light currents $J_{1,2}$ and of the 
chromo-magnetic operator, respectively. The matrix 
\begin{equation}\label{gammaA}
   \bm{\gamma_A}
   = \left( \begin{array}{ccc}
   -2(\gamma_2+\gamma_3)~ & 2\gamma_3 & ~2\gamma_2 \\
   0 & 0 & 0 \\
   -(\gamma_2+\gamma_3) & \gamma_3 & \gamma_2
   \end{array} \right)
\end{equation} 
describing the mixing of the local operators $Q_{1\dots 3}$ into 
$Q_{4\dots 6}$ has been calculated at two-loop order in \cite{Gabr}. 
Explicit expression for the two-loop anomalous dimensions will be 
presented in Section~\ref{sec:analysis}.

It was observed in \cite{MN94} that there are numerically large 
discrepancies between the one-loop and leading-logarithmic 
approximations to the Wilson coefficients $B_j(\mu)$, which could only
be resolved at NLO in RG-improved perturbation theory. The missing 
piece in the NLO analysis is the calculation of the anomalous dimension 
matrix $\bm{\gamma_C}$ describing the mixing of the bilocal operators 
$Q_{7\dots 10}$ into the local operators $Q_{4\dots 6}$. In this paper, 
we report the calculation of this matrix at two-loop order. It requires 
the evaluation of two-loop tensor integrals in the HQET, which are 
infrared (IR) singular when one of the external lines is taken on-shell. 
A general algorithm for computing such integrals has been developed in 
\cite{ABN}. For our analysis we will generalize the above discussion and 
consider an arbitrary Dirac structure $\Gamma$ of the current 
$\bar q\,\Gamma h_v$ in the time-ordered products. At the end we will 
study the particular case of vector currents in detail.

The reader interested only in the final results of our calculation but
not its technical details can continue with Section~\ref{sec:analysis}, 
where we present the final expressions for the Wilson coefficients 
$B_j(\mu)$ at NLO and briefly discuss some phenomenological 
applications.

\section{Operator mixing}

Our task is to calculate, at two-loop order, the mixing of bilocal 
operators, consisting of time-ordered products of a local current 
$\bar q\,\Gamma h_v$ with either the kinetic operator or the 
chromo-magnetic operator, into local dimension-4 current operators with 
a derivative acting on the light-quark field. Here $\Gamma$ can be an 
arbitrary Dirac matrix, depending on the Lorentz structure of the 
original current that is expanded in terms of HQET operators. 

We start by constructing, in the HQET, a basis of the relevant operators
that mix under renormalization. Since ultimately our interest is in the 
matrix elements of these operators between physical hadron states, it is 
sufficient to consider gauge-invariant operators that do not vanish by 
the equations of motions. For the case where the bilocal operator 
contains the kinetic operator, such a basis consists of the two 
operators
\begin{eqnarray}\label{Okin}
   O_1^{\rm kin} &=& \bar q\,(iv\!\cdot\!D)^\dagger \Gamma h_v \,,
    \nonumber\\
   O_T^{\rm kin} &=& i\!\int\!\mbox{d}^4x\,\mbox{T}\{ 
    \bar q\,\Gamma h_v(0),\bar h_v (iD)^2 h_v(x) \} \,.
\end{eqnarray}
If the bilocal operator contains the magnetic operator, the basis 
consists of the three operators
\begin{eqnarray}\label{Omag}
   O_1^{\rm mag} &=& -\frac14\,\bar q\,(iv\!\cdot\!D)^\dagger 
    \sigma_{\mu\nu} \Gamma\,(1+\vslash)\,\sigma^{\mu\nu} h_v \,,
    \nonumber\\
   O_2^{\rm mag} &=& -\frac14\,\bar q\,(iD_\nu)^\dagger
    i\gamma_\mu\vslash\,
    \Gamma\,(1+\vslash)\,\sigma^{\mu\nu} h_v \,, \nonumber\\
   O_T^{\rm mag} &=& i\!\int\!\mbox{d}^4x\,\mbox{T}\{
    \bar q\,\Gamma h_v(0),
    \frac{g_s}{2}\,\bar h_v\sigma_{\mu\nu} G^{\mu\nu} h_v(x) \} \,.
\end{eqnarray}
For each of the two cases, we define a matrix $\bm{Z}$ of 
renormalization constants, which absorb the ultraviolet (UV) divergences 
in the matrix elements of the bare operators, by the relation 
$O_i=\sum_j\,Z_{ij}\,O_{j,{\rm bare}}$. The matrix $\bm{\gamma}$ of the 
anomalous dimensions, which govern the scale dependence of the 
renormalized operators, is given by
\begin{equation}
   \bm{\gamma} = - \frac{\mbox{d}\bm{Z}}{\mbox{d}\ln\mu}\,\bm{Z}^{-1} \,.
\end{equation}
In a minimal subtraction scheme, the renormalization constants are
defined to remove the $1/\ep$ poles arising in the calculation of the
bare Green functions with insertions of the operators $O_i$ in
dimensional regularization, i.e.\ in $d=4-2\ep$ space--time dimensions.
Hence,
\begin{equation}
   \bm{Z} = \bm{1} + \sum_{k=0}^\infty\,\frac{1}{\ep^k}\,
   \bm{Z}^{(k)} \,.
\end{equation}
The requirement that the anomalous dimension matrix be finite in the
limit $\ep\to 0$ implies the relations \cite{Flor}
\begin{equation}\label{magic}
   \mbox{\bm{\gamma}} 
   = 2\alpha_s \frac{\partial\bm{Z}^{(1)}}{\partial\alpha_s} \,,
   \qquad
   \alpha_s \frac{\partial\bm{Z}^{(2)}}{\partial\alpha_s}
   = \alpha_s \frac{\partial\bm{Z}^{(1)}}{\partial\alpha_s}
   \left( \bm{Z}^{(1)} + \frac{\beta(\alpha_s)}{\alpha_s} \right) \,,
\end{equation}
where $\beta(\alpha_s)=\mbox{d}\alpha_s/\mbox{d}\ln\!\mu^2$ is the
$\beta$ function. The first equation shows that the anomalous dimension
matrix can be obtained from the coefficient of the $1/\ep$ pole in
$\bm{Z}$, whereas the second one implies a nontrivial constraint on
the coefficient of the $1/\ep^2$ pole, arising at two-loop and higher
order.

In our case, the structure of the two $\bm{Z}$ matrices is
\begin{equation}\label{Zmat}
   \bm{Z}^{\rm kin} = \left( \begin{array}{cc}
    ~Z_1~ & 0 \\
    Z_{T1}^{\rm kin} & ~Z_1~ 
   \end{array} \right) \,, \qquad
   \bm{Z}^{\rm mag} = \left( \begin{array}{ccc}
    ~Z_1~ & 0 & 0 \\
    ~Z_4~ & Z_1+Z_2 & 0 \\
    Z_{T1}^{\rm mag} & Z_{T2}^{\rm mag} & Z_1 Z_{\rm mag} 
   \end{array} \right) \,,
\end{equation}
where the renormalization constants $Z_1$, $Z_2$ and $Z_4$ determine the 
mixing of local, dimension-4 heavy--light currents and have been 
calculated at two-loop order in \cite{Gabr}, whereas $Z_{\rm mag}$ is 
the renormalization constant of the chromo-magnetic operator, which has 
been computed at two-loop order in \cite{ABN}. Here we will calculate 
the remaining entries $Z_{Ti}$ with the same accuracy.

Previous authors have calculated the mixing of the bilocal operators 
into local operators at the one-loop order, finding \cite{FaGr,MN94}
\begin{equation}\label{Z1loop}
   Z_{T1}^{\rm kin} = - \frac{C_F\alpha_s}{\pi\ep} + O(\alpha_s^2) 
   \,,\qquad
   Z_{T1}^{\rm mag} = Z_{T2}^{\rm mag} 
   = - \frac{C_F\alpha_s}{4\pi\ep} + O(\alpha_s^2) \,.
\end{equation}
Using the known expression for the other $Z$-factors, this information 
can be used to predict the $1/\ep^2$ poles in the two-loop coefficients
of the renormalization constants. From the second relation in 
(\ref{magic}), we obtain
\begin{eqnarray}\label{ep2terms}
   Z_{T1}^{{\rm kin}\,(2)} &=& \left( \frac{\alpha_s}{4\pi} \right)^2
    \left[ 6 C_F^2 + \frac{22}{3}\,C_F C_A - \frac{8}{3}\,C_F T_F\,n_f
    \right] + O(\alpha_s^3) \,, \nonumber\\
   Z_{T1}^{{\rm mag}\,(2)} &=& \left( \frac{\alpha_s}{4\pi} \right)^2
    \left[ \frac{7}{4}\,C_F^2 + \frac{4}{3}\,C_F C_A
    - \frac{2}{3}\,C_F T_F\,n_f \right] + O(\alpha_s^3) \,, \nonumber\\
   Z_{T2}^{{\rm mag}\,(2)} &=& \left( \frac{\alpha_s}{4\pi} \right)^2
    \left[ \frac{3}{4}\,C_F^2 + \frac{4}{3}\,C_F C_A
    - \frac{2}{3}\,C_F T_F\,n_f \right] + O(\alpha_s^3) \,.
\end{eqnarray}
Here $C_A=N$, $C_F=\frac 12 (N^2-1)/N$ and $T_F=\frac 12$ are the
color factors for an $SU(N)$ gauge group, and $n_f$ is the number of
light-quark flavors. These relations will provide a check on our
two-loop results.

\section{Two-loop calculation}

\begin{figure}[htb]
\begin{center}
\begin{tabular}{ccccccccc}
\raisebox{0.029\textwidth}{\lp{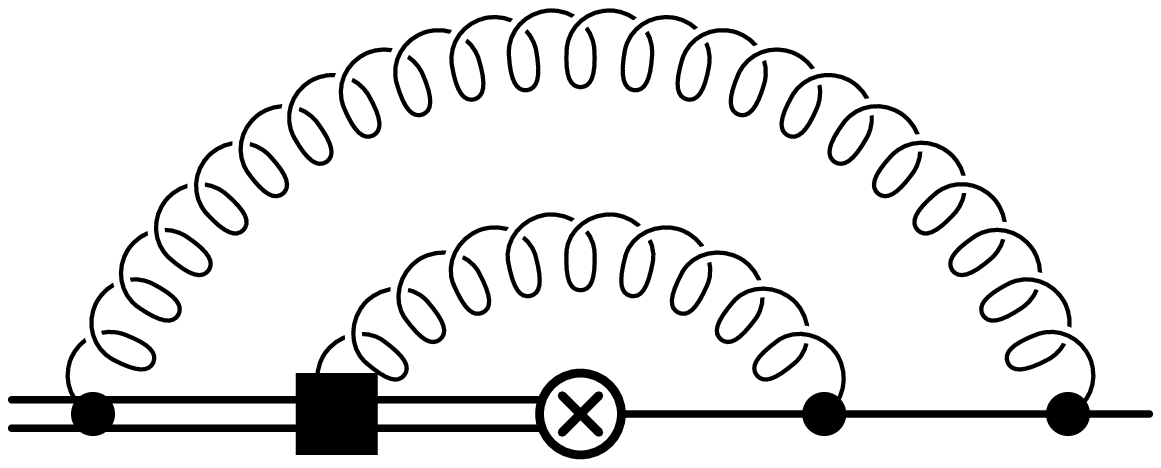}} & & \raisebox{0.029\textwidth}{\lp{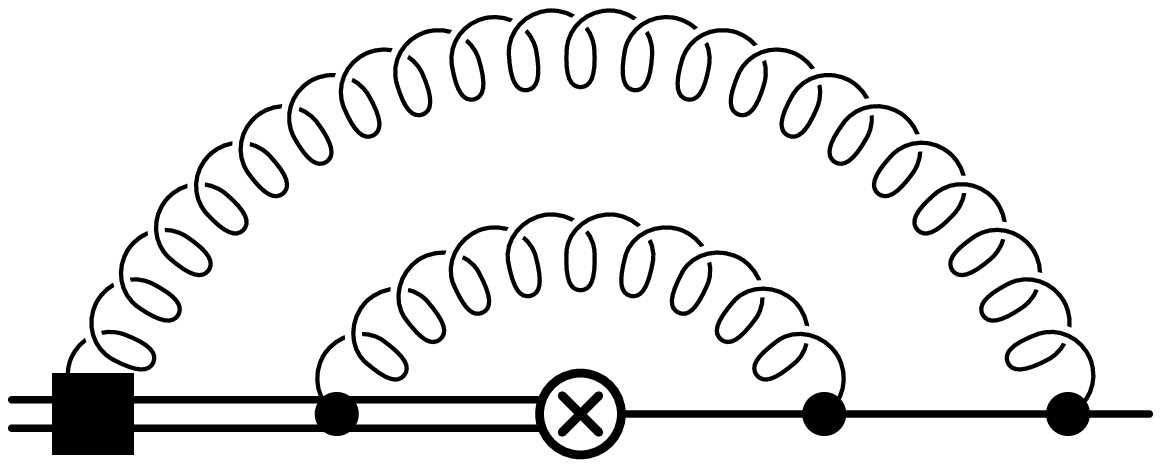}}
 & & \lp{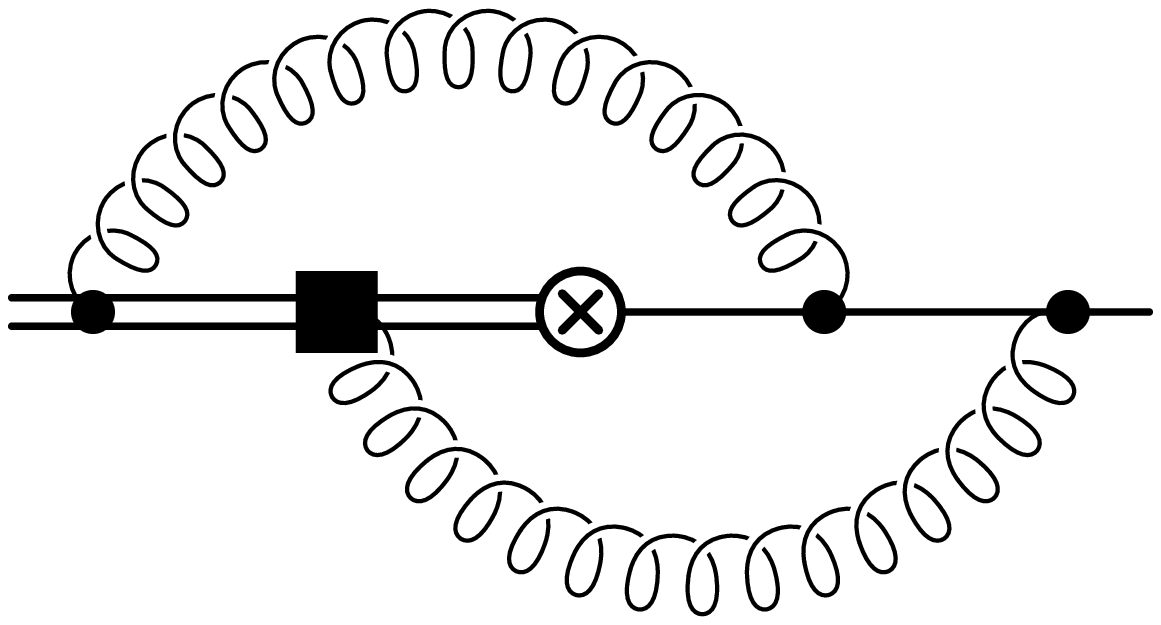} & & \lp{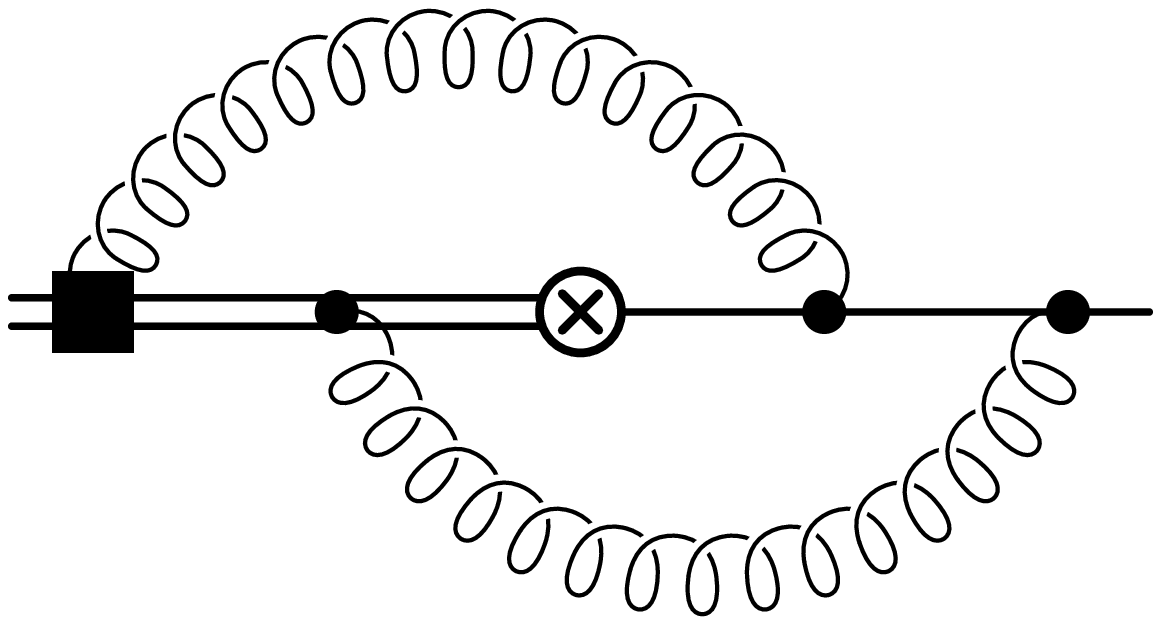} &&  \raisebox{0.010\textwidth}{\lp{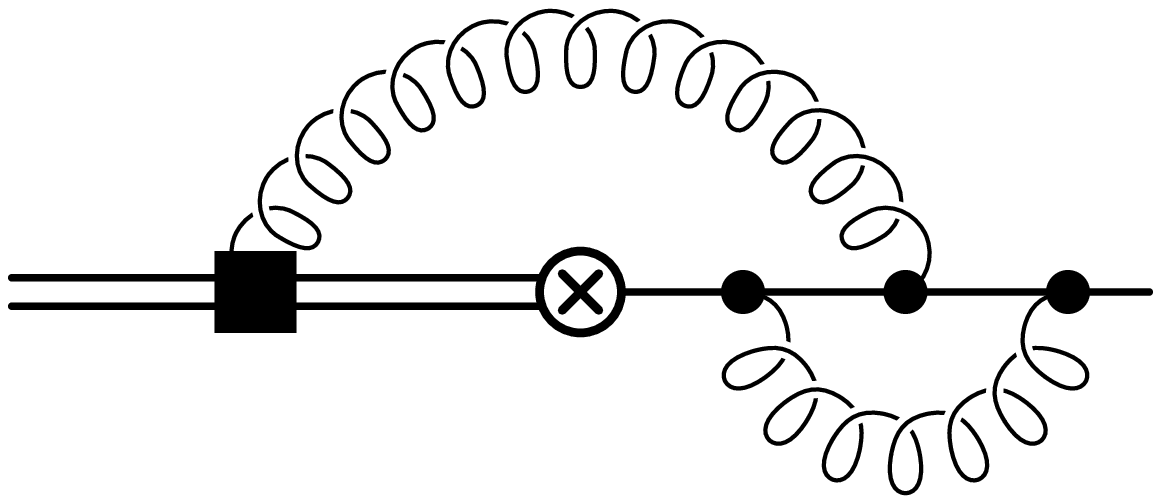}} \\ 
&&&&&&&&\\
 \lp{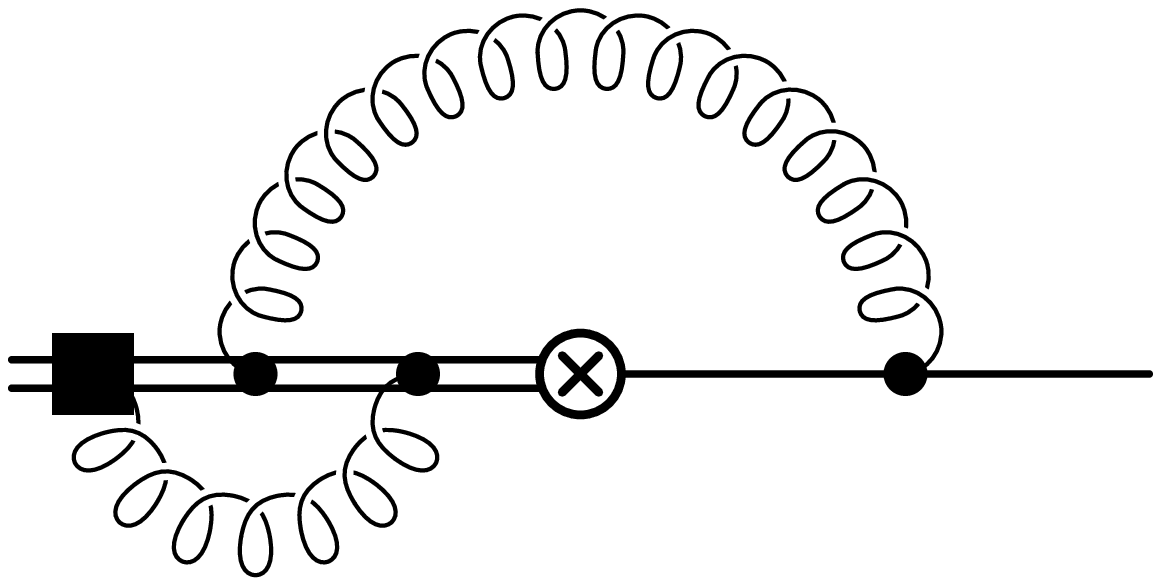} & & \lp{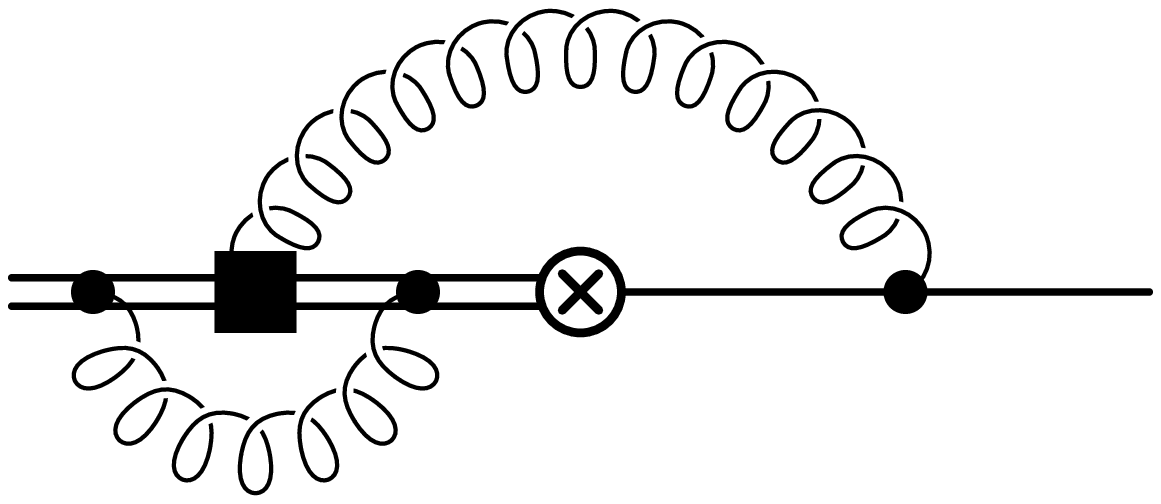} & & \lp{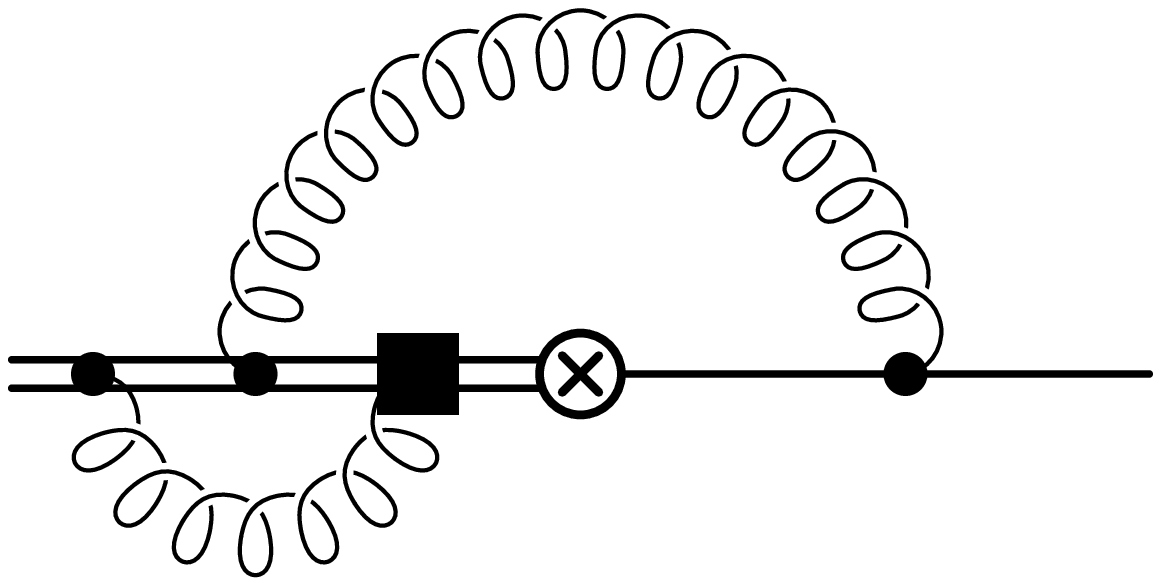} &
 &\raisebox{0.018\textwidth}{\lp{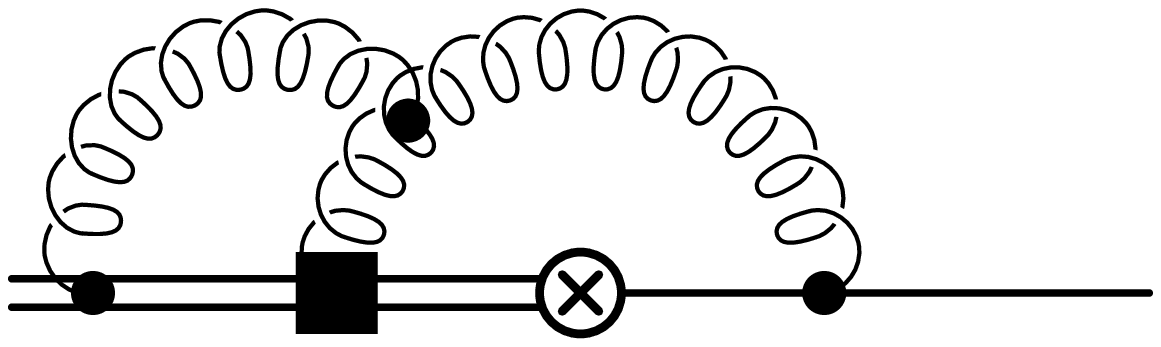}}&&\raisebox{0.018\textwidth}{\lp{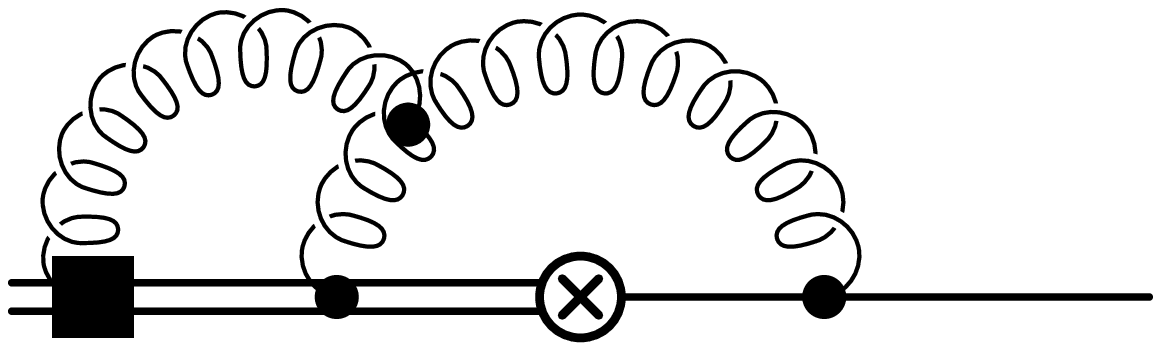}}
 \\
&&&&&&&&\\
\raisebox{0.018\textwidth}{\lp{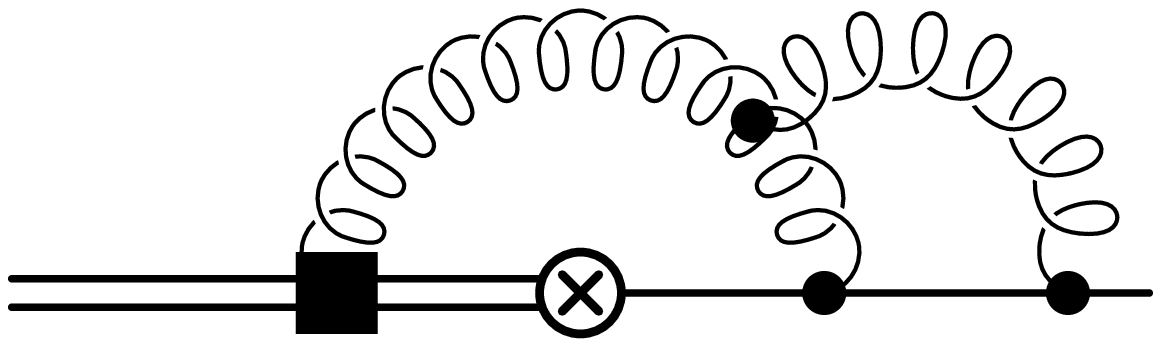}} & & \lp{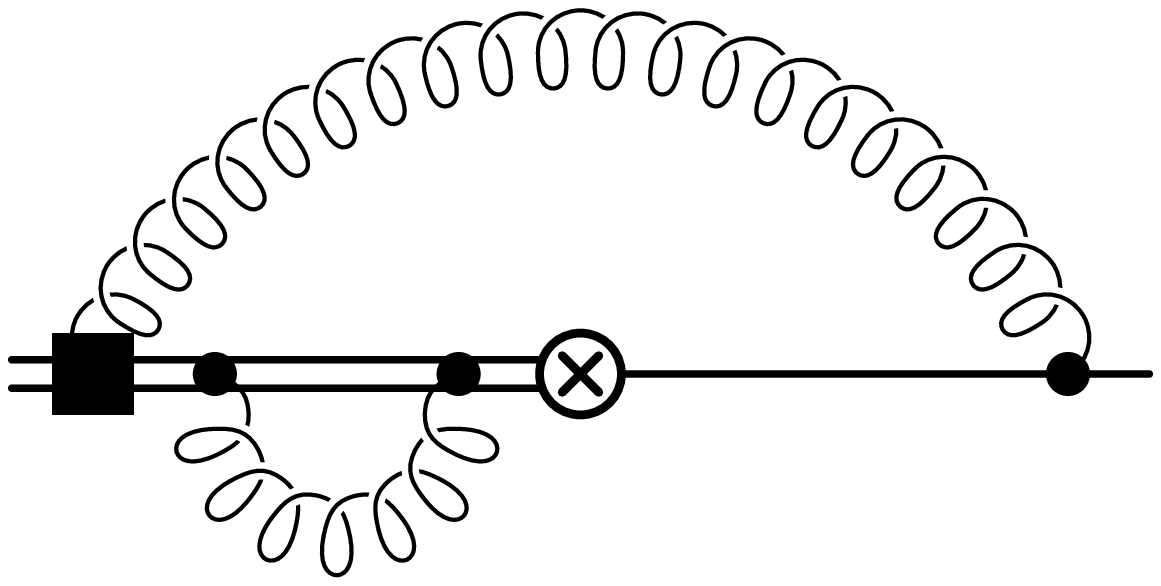} &&\lp{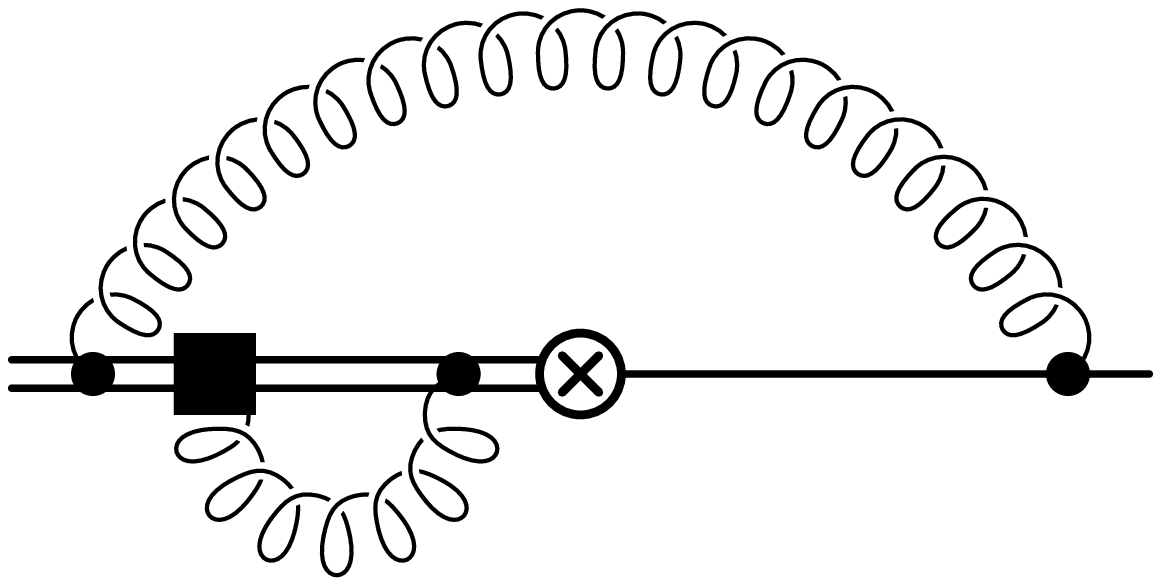}  & &  \lp{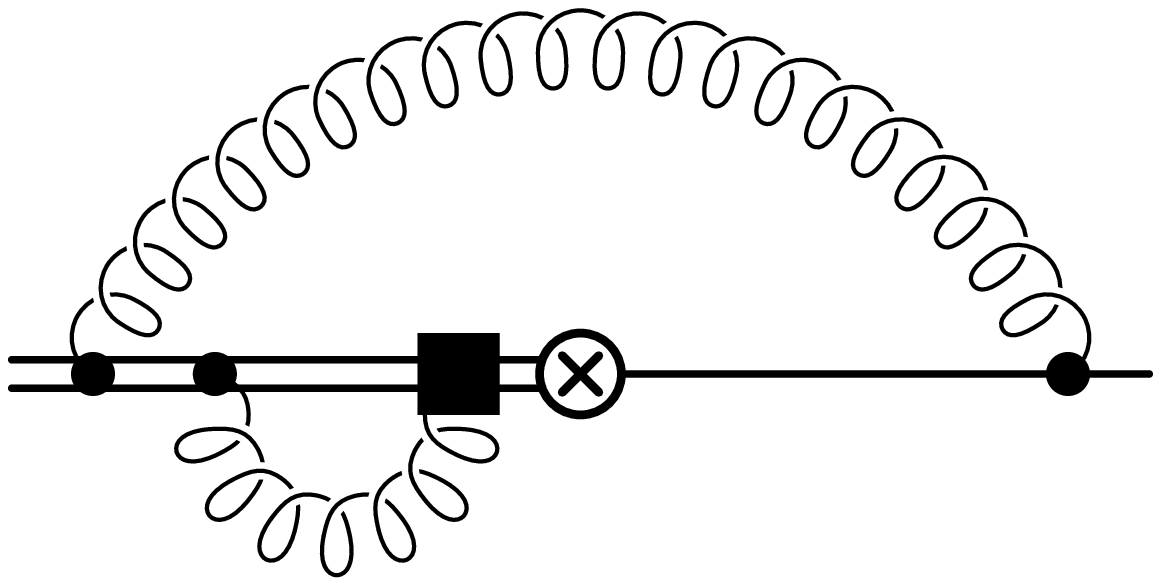}
 && \lp{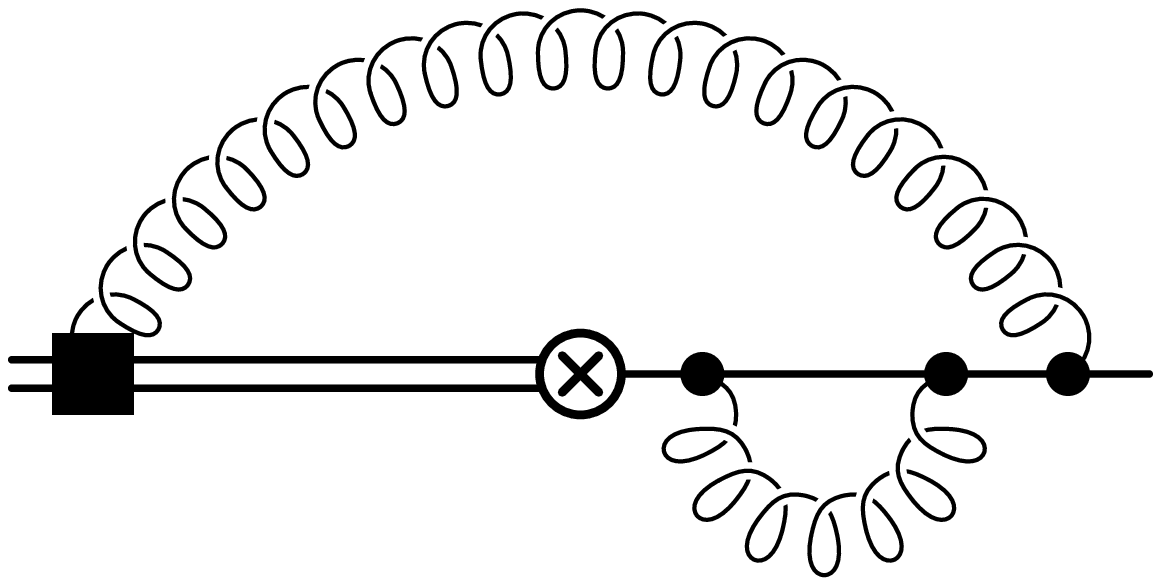}\\
 &&&&&&&&\\
\raisebox{0.029\textwidth}{\lp{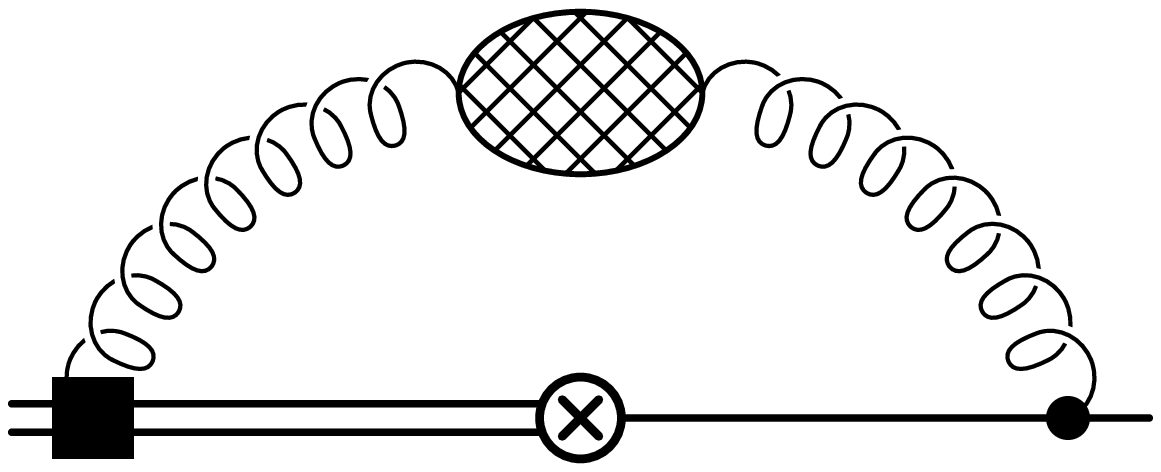}} &
 & \raisebox{0.01\textwidth}{\lp{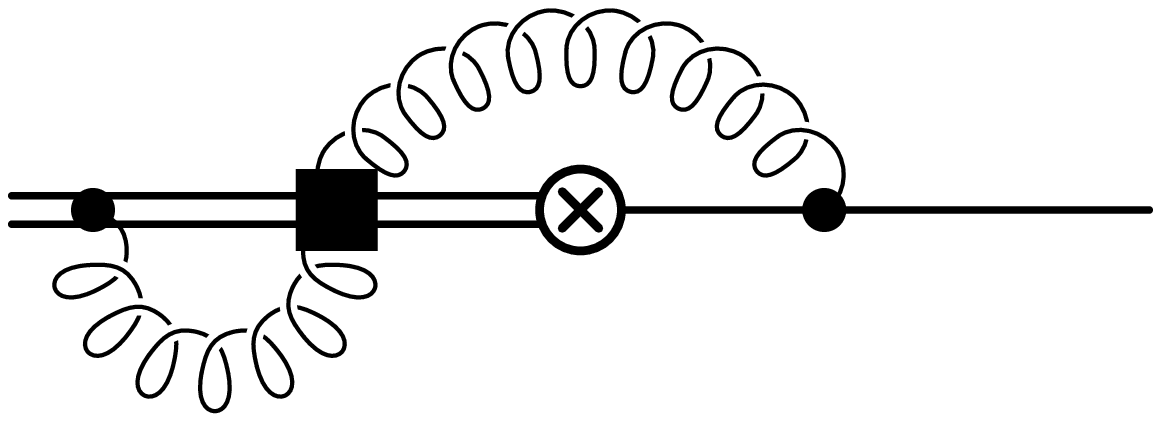}} &
 & \raisebox{0.01\textwidth}{\lp{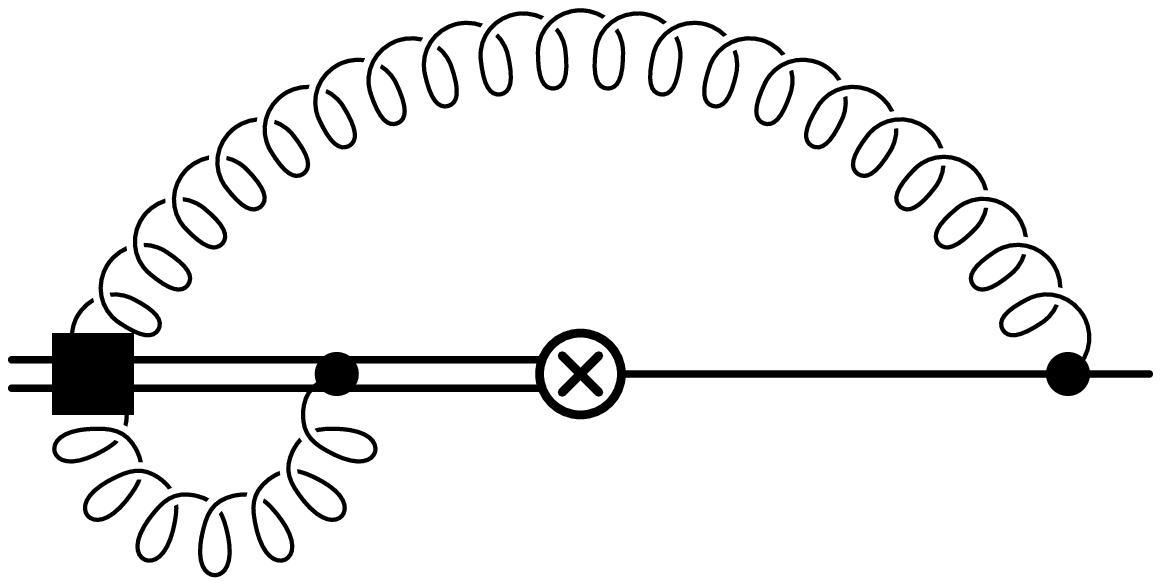}}&
 & \lp{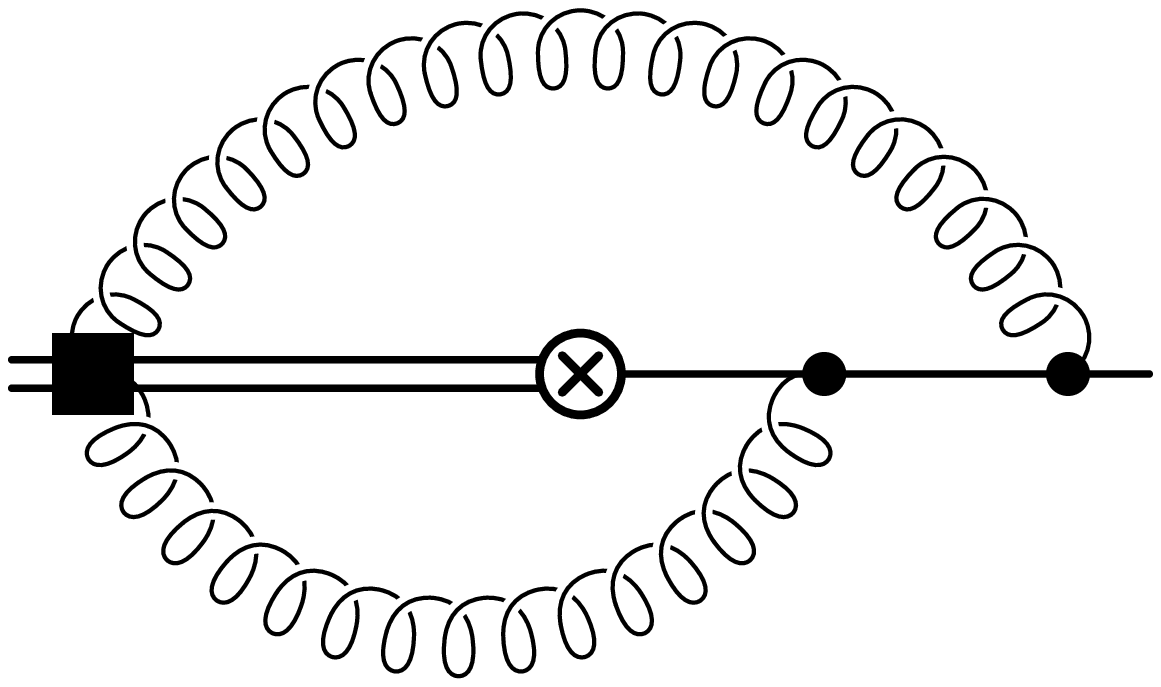} && \raisebox{0.029\textwidth}{\lp{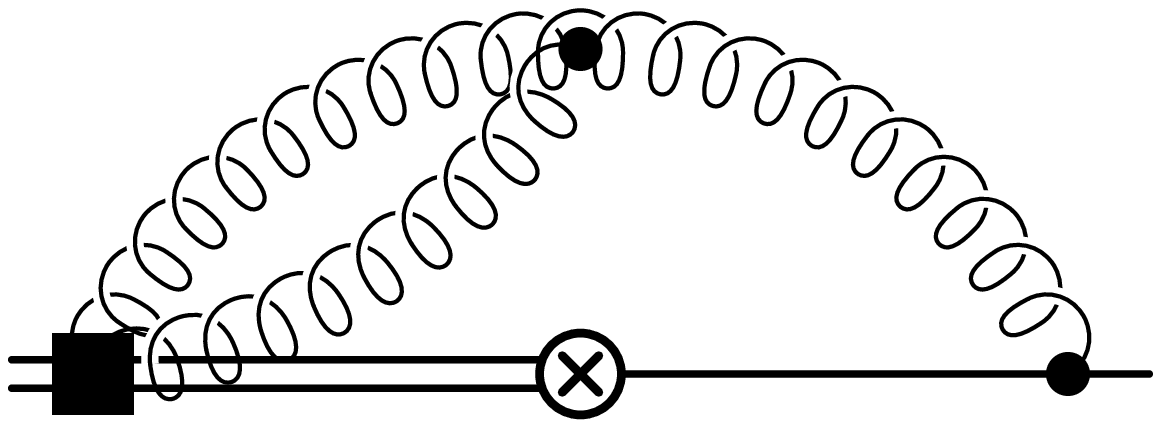}}  \\
 &&&&&&&&\\
\raisebox{0.008\textwidth}{\lp{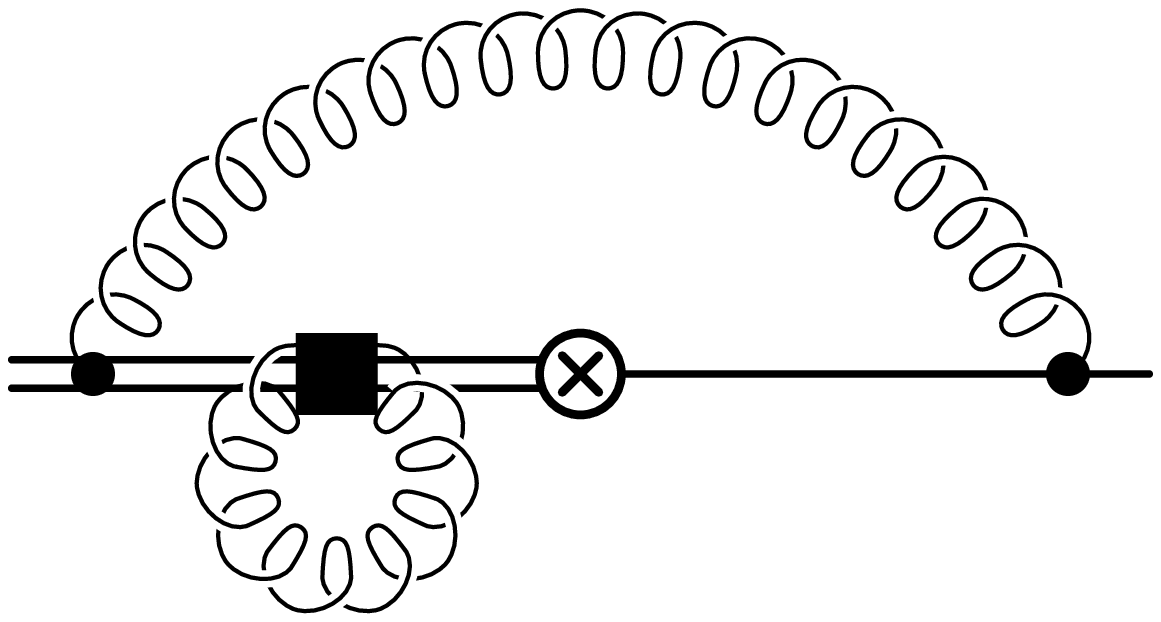}} &
 & \raisebox{0.029\textwidth}{\lp{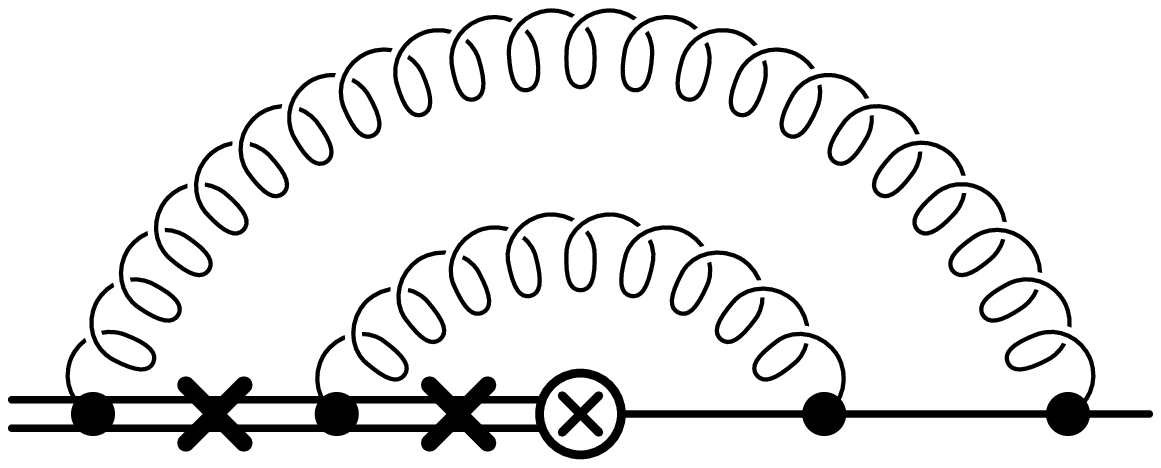}} & &\lp{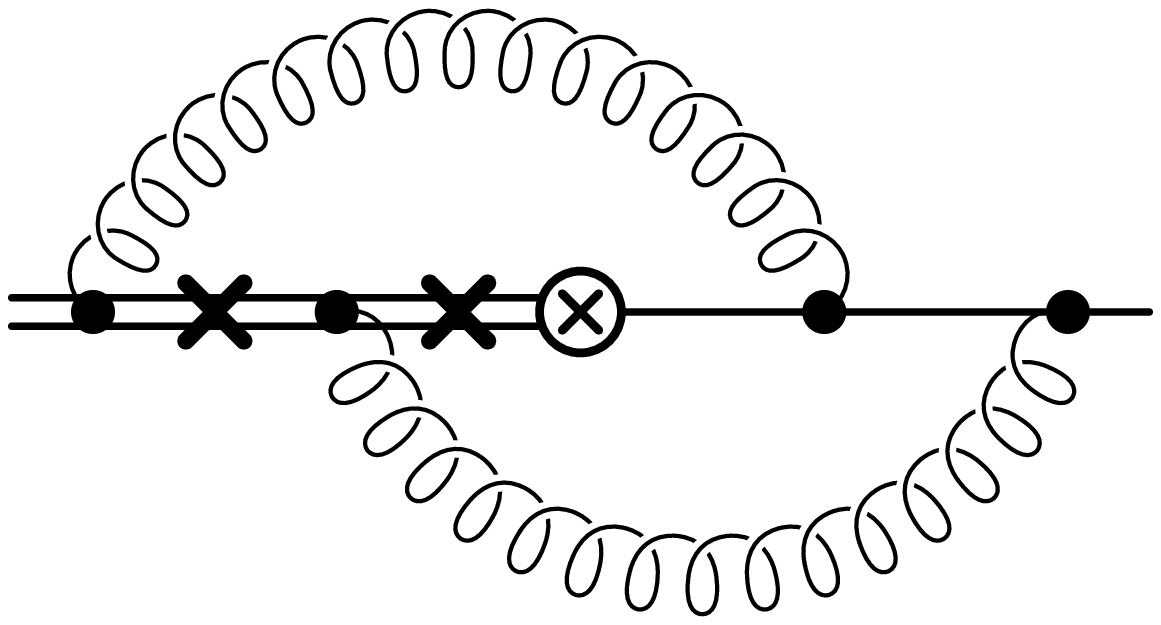} &
 &\raisebox{0.01\textwidth}{\lp{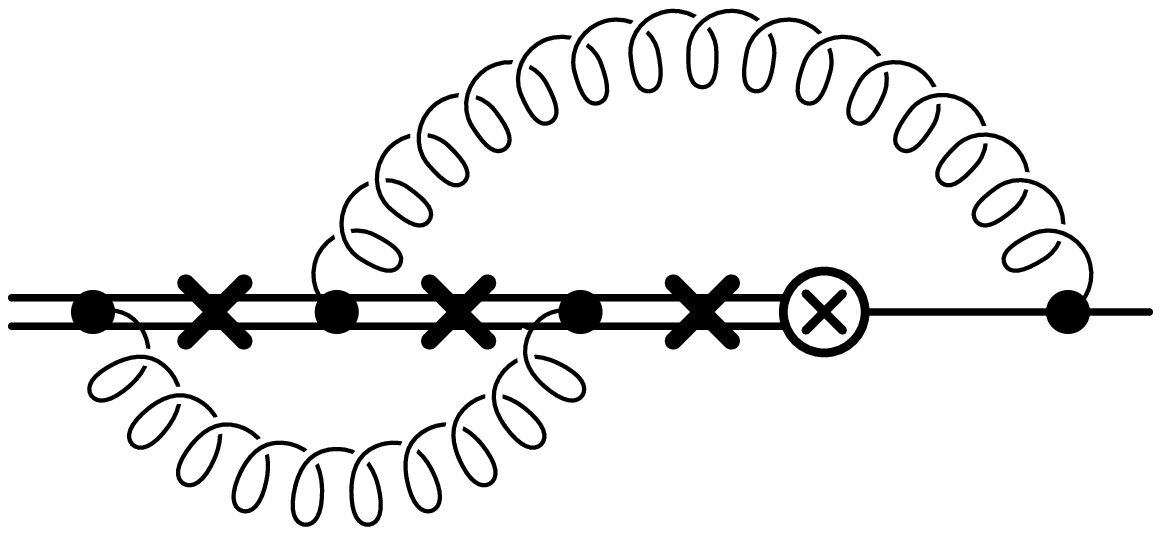}} &
 & \raisebox{0.01\textwidth}{\lp{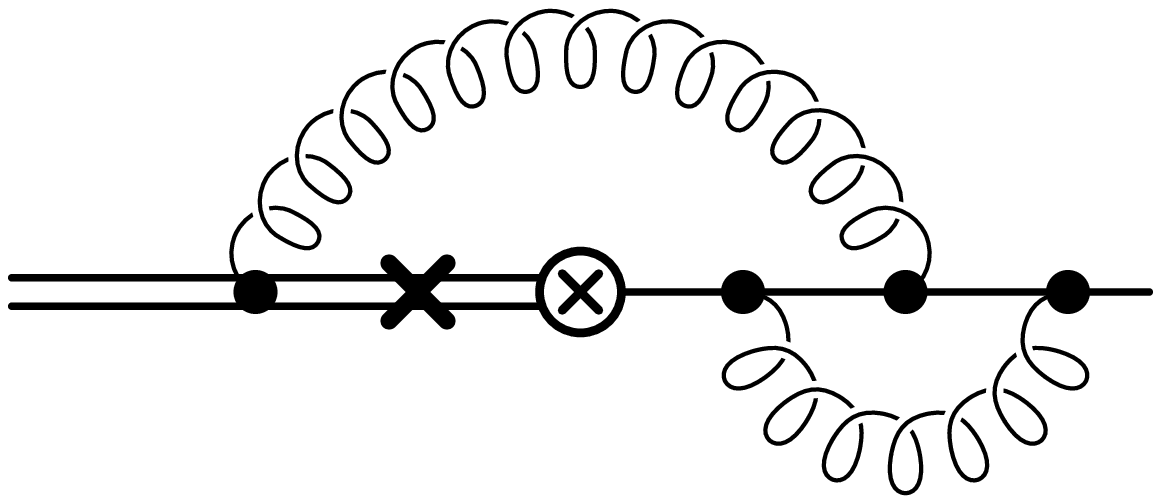}} \\
 &&&&&&&&\\
\raisebox{0.018\textwidth}{\lp{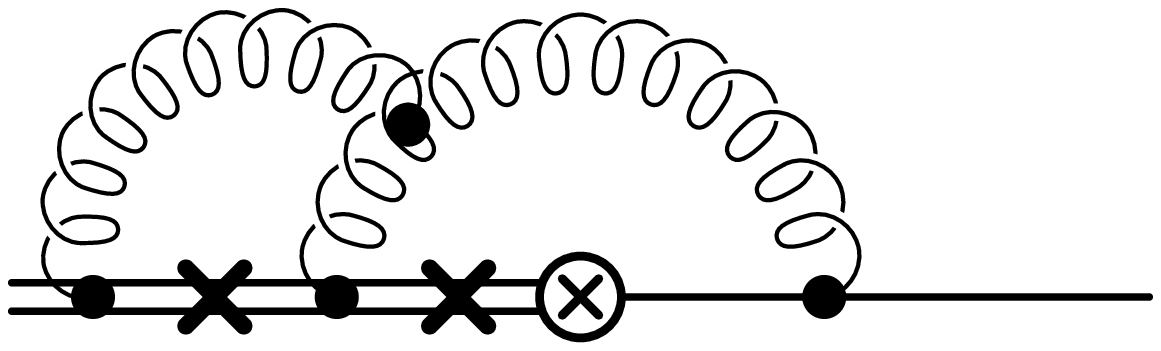}} &
 &\raisebox{0.018\textwidth}{\lp{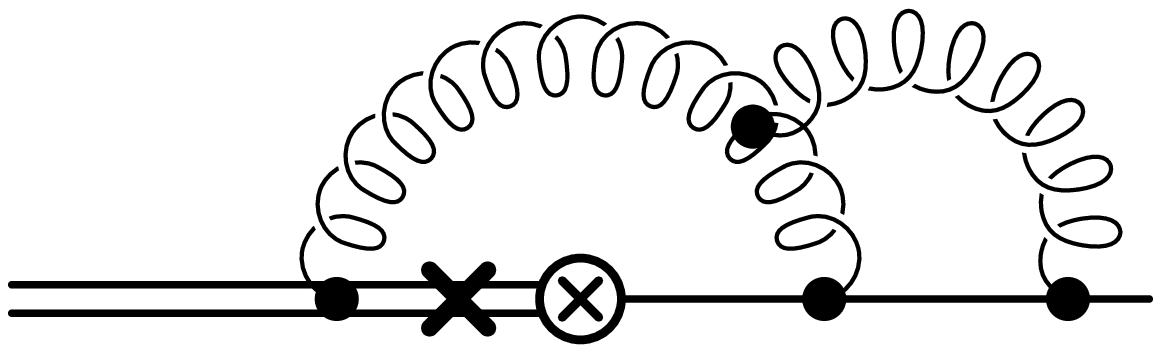}} & &\lp{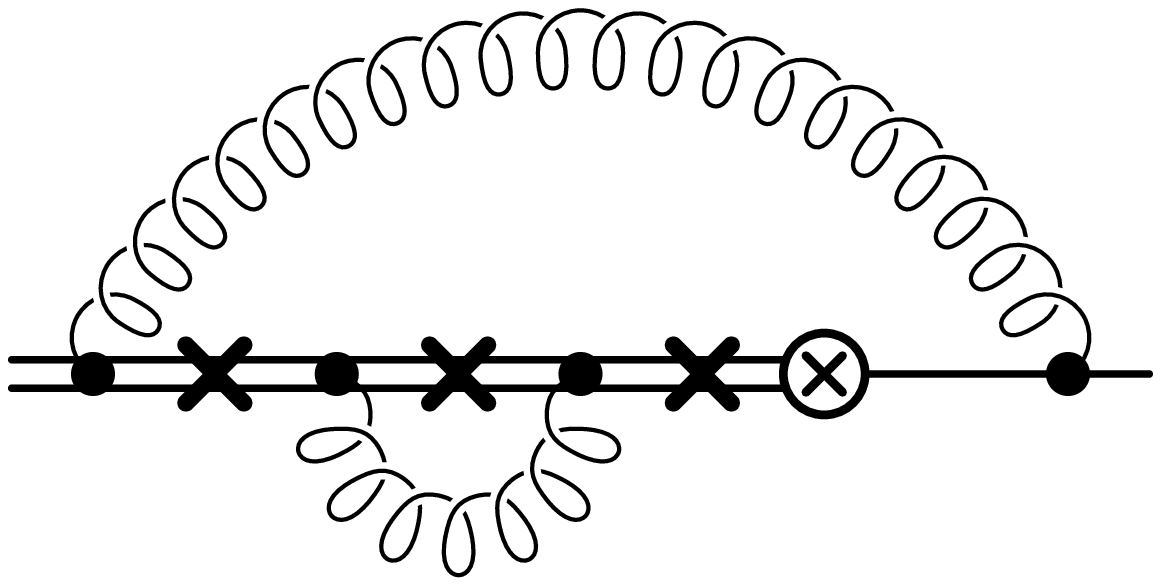}& &\lp{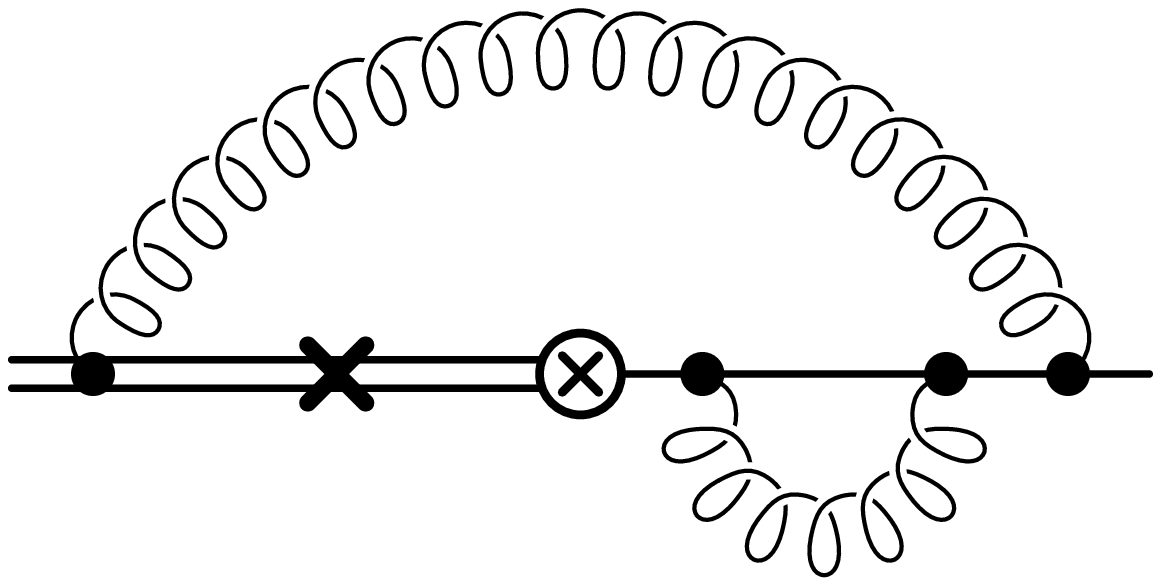} &
 & \raisebox{0.018\textwidth}{\lp{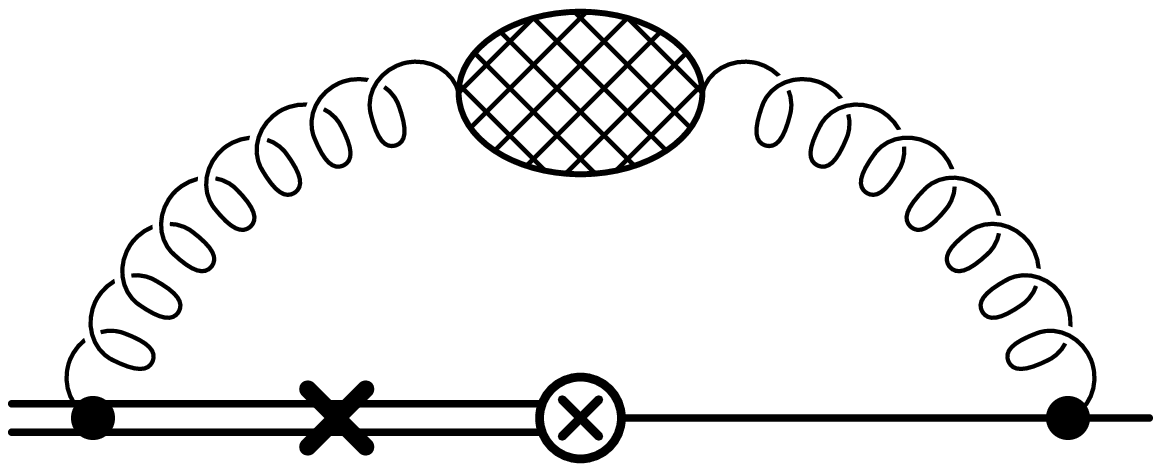}} \\
\end{tabular}
\end{center}
\vspace{-0.2cm}
\centerline{\parbox{14.5cm}{
\caption{\label{fig:2loop}
Two-loop diagrams contributing to 
the renormalization 
constants $Z_{Ti}$. The heavy quark is represented by a double line.
The crossed circles represent the leading-order current, while the black 
squares represent an insertion of the kinetic or the chromo-magnetic 
operator. The shaded blobs represent one-loop insertions of the gluon 
self-energy. The last nine diagrams contribute only to insertions of the 
kinetic operator. In these graphs, the crosses on heavy-quark lines with 
momentum $p$ mark the different positions where we insert $ip^2$.}}}
\end{figure}

To obtain the renormalization constants $Z_{Ti}$ at order $\alpha_s^2$, 
we calculate the insertions of the bilocal operators $O_T^{\rm kin}$ and 
$O_T^{\rm mag}$ into the amputated Green function with a heavy and a 
light quark to two-loop order. The relevant diagrams are shown in 
Figure~\ref{fig:2loop}. We use dimensional regularization in $d=4-2\ep$
space--time dimensions and work in the $\overline{\mbox{MS}}$ subtraction 
scheme. The relation between the renormalized and the bare, regularized 
amputated Green functions is
\begin{equation}\label{Gamrel}
   \Gamma_{{\rm ren},i}(v,p,\dots)
   = (Z_h Z_q)^{1/2} Z_{ij}\,\Gamma_{{\rm reg},j}(v,p,\dots;\ep)
   = \mbox{finite}. 
\end{equation}
To compute the mixing of the bilocal operators into the local operators 
with a derivative acting on the light-quark field, we only need to keep 
terms linear in the momentum $p$ of the light quark. Because the pole 
parts are polynomial in the external momenta, we can first take a 
derivative with respect to $p$ and then set $p=0$, so that all integrals 
are of propagator type and depend on the single variable 
$\omega=v\cdot k$. However, this method fails for some of the diagrams, 
for which setting $p=0$ after differentiation leads to infrared (IR) 
divergences. In these cases, we apply a variant of the so-called $R^*$
operation \cite{ABN,Che}, which compensates these IR poles by a 
recursive construction of counterterms for the IR-divergent subgraphs.

Consider, as an example, the first diagram in Figure~\ref{fig:2loop}.
Its contribution is proportional to the integral
\begin{equation}
   D_1 = \int\!{\rm d}^ds\,{\rm d}^dt\,
   \frac{(s-t)_\alpha\,(t+p)_\beta\,(s+p)_\gamma}
        {t^2 (t+p)^2 (s+p)^2 (s-t)^2
         (v\!\cdot\!s+\omega)(v\!\cdot\!t+\omega)} \,.
\end{equation}
When linearizing this expression in $p$, we encounter an IR divergence
from the region $t\to 0$, which can be removed by adding and 
subtracting the IR subtraction term
\begin{equation}
   D_1^{\rm IR} = \int\!{\rm d}^ds\,
   \frac{s_\alpha\,s_\gamma}
        {\big(s^2\big)^2(v\!\cdot\!s+\omega)\,\omega}
   \int\!{\rm d}^dt\,
   \frac{(t+p)_\beta}{t^2 (t+p)^2} \,.
\end{equation}
The difference $D_1-D_1^{\rm IR}$ can be evaluated using a naive
linearization in $p$, since the behavior of the IR subtraction term
for $t\to 0$ is the same as that of the original integral. However,
because the expansion of $D_1^{\rm IR}$ involves tadpole integrals that
vanish in dimensional regularization, the difference $(D_1-D_1^{\rm
IR})_{\rm linearized}$ coincides with the naive linearization of the
original expression for $D_1$. The contribution $D_1^{\rm IR}$, which
is necessary to subtract the IR subdivergence of the original diagram,
factorizes into an IR counterterm
\begin{equation}
   \int\!{\rm d}^d t\,
   \frac{\gamma^\beta(t+p)_\beta}{t^2 (t+p)^2}
   = i\pi^{d/2}\,\pslash\,(-p^2)^{-\ep}\,
   \frac{\Gamma(d/2)\,\Gamma(d/2-1)\,\Gamma(2-d/2)}{\Gamma(d-1)} \,.
\end{equation}
and the original diagram with the lines of the IR-sensitive subgraph 
removed, and with $t$ and $p$ set to zero.

The remaining two-loop tensor integrals are of the general form
\begin{eqnarray}
   &&\int\mbox{d}^ds\,\mbox{d}^dt\,
   \left( \frac{\omega}{v\!\cdot\!s+\omega} \right)^{\alpha_1}
   \left( \frac{\omega}{v\!\cdot\!t+\omega} \right)^{\alpha_2}
   \frac{s_{\mu_1}\dots s_{\mu_n} t^{\nu_1}\dots t^{\nu_m}}
    {\big(-s^2\big)^{\alpha_3} \big(-t^2\big)^{\alpha_4}
     \big[-(s-t)^2\big]^{\alpha_5}}
    \nonumber\\
   &&\hspace{2.2cm} \equiv
    -\pi^d (-2\omega)^{2(d-\alpha_3-\alpha_4-\alpha_5)+n+m}\,
    I_{\mu_1\dots\mu_n}^{\nu_1\dots\nu_m}(v;\{\alpha_i\}) \,.
\end{eqnarray}
Using the method of integration by parts \cite{BrGr,Chet}, one obtains
the recurrence relation \cite{ABN}
\begin{eqnarray}
   &&\left[ (d-\alpha_1-\alpha_3-2\alpha_5+n)
    + \alpha_3\,\bm{3^+} (\bm{4^-} - \bm{5^-})
    + \alpha_1\,\bm{1^+ 2^-} \right]\,
    I_{\mu_1\dots\mu_n}^{\nu_1\dots\nu_m}(v;\{\alpha_i\}) \nonumber\\
   &&\hspace{2.2cm} = \sum_{j=1}^n\,
    I_{\mu_1\dots[\mu_j]\dots\mu_n}^{\mu_j\nu_1\dots\nu_m}
    (v;\{\alpha_i\}) \,,
\end{eqnarray}
which can be used to express any two-loop integral in terms of degenerate
integrals, which have $\alpha_2=0$, $\alpha_4=0$ or $\alpha_5=0$. Here
$\bm{1^+}$ is an operator raising the index $\alpha_1$ by one unit
etc., and $[\mu_j]$ means that this index is omitted. The degenerate
integrals can be related in a straightforward way to products of
one-loop tensor integrals \cite{ABN}. 

Using this technique, we have calculated the pole parts of the two-loop 
diagrams in the `t~Hooft--Feynman gauge. The result is a linear 
combination of terms with the Dirac structures corresponding to the 
local operators $O_1^{\rm kin}$ and $O_{1,2}^{\rm mag}$ defined in 
(\ref{Okin}) and (\ref{Omag}). The results are summarized in the first 
two columns of Table~\ref{tab:2loop}. In the $\overline{\mbox{MS}}$ 
scheme, the renormalization scale $\mu$ is introduced by the replacement 
of the bare coupling constant with the renormalized one through the 
relation $g_s^{\rm bare}=\tilde\mu^\epsilon Z_g\,g_s$ with 
$\tilde\mu=\mu\,e^{\gamma_E/2} (4\pi)^{-1/2}$.

\begin{table}[t]
\caption{\label{tab:2loop}
Two-loop and counterterm contributions in units of $(\alpha_s/4\pi)^2$}
\begin{center}
\begin{tabular}{|l|l|c|c|c|c|}\hline\hline
Structure & Color
 & \rule[-0.35cm]{0cm}{0.95cm}
  $\left( \frac{-2\omega}{\mu} \right)^{-4\ep}$
  \rule[-0.35cm]{0cm}{0.95cm}
 & $\!\left( \frac{-2\omega}{\mu} \right)^{-2\ep}\!
    \left( \frac{-p^2}{\mu^2} \right)^{-\ep}\!\!$
 & $\left( \frac{-2\omega}{\mu} \right)^{-2\ep}$
 & $\left( \frac{-p^2}{\mu^2} \right)^{-\ep}$ \\
\hline
$O_1^{\rm kin}$ \rule{0cm}{0.5cm} 
& $C_F^2$ &
 $\frac{4}{\ep^2}
  +\left(-\frac{4}{3}+\frac{8\pi^2}{9}\right)\!\frac{1}{\ep}$ &
 $-\frac{4}{\ep}$ &
 $-\frac{8}{\ep^2}+\frac{8}{\ep}$ &
 0 \\[0.15cm]
& $C_F C_A$ & 
 $\frac{53}{6\ep^2}
  +\left(-\frac{287}{18}-\frac{4\pi^2}{3}\right)\!\frac{1}{\ep}$ &
 $-\frac{3}{2\ep^2}-\frac{1}{\ep}$ &
 $-\frac{97}{6\ep^2}+\frac{185}{6\ep}$ &
 $\frac{3}{2\ep^2}+\frac{3}{\ep}$ \\[0.15cm]
& $C_F T_F\,n_f$ &
 $-\frac{8}{3\ep^2}+\frac{56}{9\ep}$ &
 0 &
 $\frac{16}{3\ep^2}-\frac{32}{3\ep}$ &
 0 \\[0.15cm]
\hline
$O_1^{\rm mag}$ \rule{0cm}{0.5cm}
& $C_F^2$ &
 $\frac{9}{4\ep^2}
  +\left(\frac{13}{3}+\frac{10\pi^2}{9}\right)\!\frac{1}{\ep}$ &
 $-\frac{1}{\ep^2}-\frac{6}{\ep}$ &
 $-\frac{7}{2\ep^2}+\frac{1}{\ep}$ &
 $\frac{1}{\ep^2}+\frac{2}{\ep}$ \\[0.15cm]
& $C_F C_A$ &
 $\frac{4}{3\ep^2}
  +\left(\frac{25}{18}-\frac{5\pi^2}{18}\right)\!\frac{1}{\ep}$ &
 0 &
 $-\frac{8}{3\ep^2}$ &
 0 \\[0.15cm]
& $C_F T_F\,n_f$ & 
 $-\frac{2}{3\ep^2}-\frac{1}{9\ep}$ &
 0 &
 $\frac{4}{3\ep^2}$ &
 0 \\[0.15cm]
\hline
$O_2^{\rm mag}$ \rule{0cm}{0.5cm} 
& $C_F^2$ &
 $\frac{5}{4\ep^2}
  +\left(-\frac{11}{6}+\frac{2\pi^2}{9}\right)\!\frac{1}{\ep}$ &
 $-\frac{1}{\ep^2}-\frac{2}{\ep}$ &
 $-\frac{3}{2\ep^2}$ &
 $\frac{1}{\ep^2}+\frac{2}{\ep}$ \\[0.15cm]
& $C_F C_A$ &
 $\frac{4}{3\ep^2}
  +\left(\frac{55}{18}-\frac{\pi^2}{18}\right)\!\frac{1}{\ep}$ &
 0 &
 $-\frac{8}{3\ep^2}$ &
 0 \\[0.15cm]
& $C_F T_F\,n_f$
 & $-\frac{2}{3\ep^2}-\frac{1}{9\ep}$ &
 0 &
 $\frac{4}{3\ep^2}$ &
 0 \\[0.15cm]
\hline\hline
\end{tabular}
\end{center}
\end{table}

The two-loop diagrams in Figure~\ref{fig:2loop} contain subdivergences,
which must be subtracted by UV counterterms. In addition to the standard
one-loop counterterms for the quark and gluon propagators and vertices,
local operator counterterms are required. To find these, we have 
calculated at one-loop order all insertions of the operators 
$O_T^{\rm kin}$ and $O_T^{\rm mag}$ into the amputated Green functions 
with a non-negative degree of divergence. In our case, those are the 
two- and three-point functions with field content $\bar q h_v$, 
$\bar q A h_v$, $\bar h_v h_v$, and $\bar h_v A h_v$. We find that in 
the 't~Hooft--Feynman gauge the UV divergences of these functions are 
removed by the counterterms
\begin{eqnarray}
   {\cal L}_{\rm c.t.}^{\rm kin} 
   &=& - \frac{C_F\alpha_s}{4\pi\ep} \bigg[
    4O^{\rm kin} - O_T^{\rm kin} - 2\bar q\,\Gamma\,iv\!\cdot\!D\,h_v
    - \bar q\,(i\Dslash)^\dagger \vslash\,\Gamma h_v \nonumber\\
   &&\hspace{1.5cm}\mbox{}+
    i\!\int\!\mbox{d}^4x\,{\rm T}\{ \bar q\,\Gamma h_v(0),
    \bar h_v (iv\!\cdot\!D)^2 h_v(x) \} \bigg] \nonumber\\
   &&\mbox{}- \frac{C_A\alpha_s}{4\pi\ep} \bigg[
    - \frac{3}{2}\,\bar q\,\Gamma\,v\!\cdot\!A\,h_v
    - \frac{3}{2}\,i\!\int\!\mbox{d}^4x\,{\rm T}\{ \bar q\,\Gamma h_v(0),
    \bar h_v\,\{iv\!\cdot\!D,v\!\cdot\!A\} h_v(x) \} \nonumber\\
   &&\hspace{1.5cm}\mbox{}+
    i\!\int\!\mbox{d}^4x\,{\rm T}\{ \bar q\,\Gamma h_v(0),
    \bar h_v\,\{iD_\mu,A^\mu\} h_v(x) \} \bigg] \,,
\end{eqnarray}
and
\begin{equation}
   {\cal L}_{\rm c.t.}^{\rm mag} 
   = - \frac{C_F\alpha_s}{4\pi\ep}
   \Big[ O_1^{\rm mag} + O_2^{\rm mag} - O_T^{\rm mag} 
   + \bar q\sigma_{\mu\nu}\Gamma P_+\,\sigma^{\mu\nu}
     iv\!\cdot\!D\,h_v \Big] \,.
\end{equation}
Since the two-loop calculation was performed off-shell, some
gauge-dependent operators must be included in addition to the
operators $O_i$ and $O_T$, as well as operators that vanish by the 
equations of motion. (The local operators vanishing by the equations of 
motion actually yield no contribution and could be dropped; however, 
one must keep these operators when they appear inside time-ordered 
products.) The results for the counterterm contributions are summarized 
in the last two columns of Table~\ref{tab:2loop}. When these 
contributions are added to the result for the sum of the two-loop 
diagrams, all nonlocal $1/\epsilon$ divergences proportional to 
$\ln(-2\omega/\mu)$ and $\ln(-p^2/\mu^2)$ cancel. This is a  
check of our calculation.

It follows from (\ref{Gamrel}) that the sum of the two-loop diagrams 
plus their counterterms determines the two-loop coefficients of the 
products $-(Z_h Z_q)^{1/2} Z_{Ti}$. To obtain the results for the 
renormalization constants $Z_{Ti}$ at order $\alpha_s^2$, we have to 
account for one-loop operator mixing and wave-function renormalization 
of the external quark fields. Our final expressions are
\begin{eqnarray}\label{Zres}
   Z_{T1}^{\rm kin} &=& \frac{C_F\alpha_s}{\pi\ep} \Bigg\{
    -4 + \frac{\alpha_s}{4\pi} \Bigg[
    C_F \left( \frac{6}{\ep} - \frac{8}{3} - \frac{8\pi^2}{9} \right) 
    + C_A \left( \frac{22}{3\ep}
    - \frac{152}{9} + \frac{4\pi^2}{3} \right) \nonumber\\
   &&\hspace{3.3cm}\mbox{}+ T_F\,n_f
    \left( - \frac{8}{3\ep} + \frac{40}{9} \right) \Bigg] \Bigg\} \,,
    \nonumber\\
   Z_{T1}^{\rm mag} &=& \frac{C_F\alpha_s}{4\pi\ep} \Bigg\{
    - 1 + \frac{\alpha_s}{4\pi} \Bigg[
    C_F \left( \frac{7}{4\ep} - \frac{4}{3} - \frac{10\pi^2}{9} \right) 
    + C_A \left( \frac{4}{3\ep}
    - \frac{25}{8} + \frac{5\pi^2}{18} \right) \nonumber\\
   &&\hspace{3.3cm}\mbox{}+ T_F\,n_f
    \left( - \frac{2}{3\ep} + \frac{1}{9} \right) \Bigg] \Bigg\} \,,
    \nonumber\\
   Z_{T2}^{\rm mag} &=& \frac{C_F\alpha_s}{4\pi\ep} \Bigg\{
    - 1 + \frac{\alpha_s}{4\pi} \Bigg[
    C_F \left( \frac{3}{4\ep} + \frac{11}{6} - \frac{2\pi^2}{9} \right)
    + C_A \left( \frac{4}{3\ep}
    - \frac{55}{18} + \frac{\pi^2}{18} \right) \nonumber\\
   &&\hspace{3.3cm}\mbox{}+ T_F\,n_f
    \left( - \frac{2}{3\ep} + \frac{1}{9} \right) \Bigg] \Bigg\} \,.
\end{eqnarray}
They are the main result of this work. Note that the $1/\ep^2$ poles
in these expressions agree with (\ref{ep2terms}). This a nontrivial 
check of our calculation.

\section{Nontrivial basis transformations}

The results derived in this work are sufficient to calculate, at the
two-loop order, the operator mixing of bilocal operators into local 
dimension-4 operators for any choice of the Dirac structure $\Gamma$. 
However, care must be taken when using the results in (\ref{Zres}) for 
a particular choice of $\Gamma$, if instead of the operators in 
(\ref{Okin}) and (\ref{Omag}) another set of basis operators is employed. 
Whereas the transformation between one operator basis and another is 
trivial at the one-loop order, it can be subtle at NLO, because in 
dimensional regularization the relations between the operators of 
different bases may depend on $\epsilon$.

Consider the general case where the operators $\{O_i\}$ are expressed 
in terms of some other operators $\{Q_j\}$ by a linear transformation 
of the form $O_i=\sum_j\,R_{ij}(\ep)\,Q_j$. Depending on the choice of 
$\Gamma$, some of the operators $O_i$ may not be independent, so the 
new set may contain fewer operators than the original one. We define a 
left-inverse matrix $\bm{L}(\ep)$ such that 
$\bm{L}(\ep) \bm{R}(\ep)=\bm{1}$. It then follows that the matrix 
$\bm{\tilde Z}$ of renormalization constants in the new basis (denoted 
by a tilde) must be such that\footnote{The corresponding relation~(19) 
of \protect\cite{Gabr} is incorrect; however, this did not affect the 
results for the anomalous dimensions obtained in that paper.}
\begin{equation}
   \bm{L}\bm{Z}\bm{R}\,\bm{\tilde Z}^{-1} \Big|_{\rm poles} = 0 \,.
\end{equation}
Next, we expand the transformation matrices as 
$\bm{R}(\ep)=\sum_n \bm{R_n}(\ep)\,\ep^n$ and
$\bm{L}(\ep)=\sum_n \bm{L_n}(\ep)\,\ep^n$ with $n\ge 0$, and extract 
from this result the coefficient of the $1/\ep$ pole in the matrix 
$\bm{\tilde Z}$, which determines the anomalous dimension matrix in the 
new basis. We find
\begin{eqnarray}
   \bm{\tilde Z}^{(1)} 
   &=& \bm{L_0} \bm{Z}^{(1)} \bm{R_0}
    + \Big[ \bm{L_1} \bm{Z}^{(2)} \bm{R_0} 
    + \bm{L_0} \bm{Z}^{(2)} \bm{R_1} \Big] \nonumber\\
   &&- \Big[ \bm{L_1} \bm{Z}^{(1)} \bm{R_0} 
    + \bm{L_0} \bm{Z}^{(1)} \bm{R_1} \Big] 
    \bm{L_0} \bm{Z}^{(1)} \bm{R_0} + \dots \,,
\end{eqnarray}
where the dots represent terms that do not contribute at two-loop order.

Armed with this general result, we now discuss the case of the vector 
current considered in the Introduction. Then the relevant Dirac 
structures are $\Gamma=\gamma^\alpha$ and $\Gamma=v^\alpha$, and the 
operators $O_i$ are conveniently expressed in terms of the operators 
$Q_{4\dots 6}$ defined in (\ref{vbasis}). For $\Gamma=\gamma^\alpha$ we 
have $O_1^{\rm kin}=Q_4$, $O_T^{\rm kin}=Q_7$, $O_T^{\rm mag}=Q_9$, and
\begin{eqnarray}\label{rel1}
   O_1^{\rm mag} &=& (1-\ep)(1+2\ep)Q_4 - 4(1-\ep)Q_5 \,, \nonumber\\
   O_2^{\rm mag} &=& \ep\,Q_4 - 2(1-\ep)Q_5 + (1-2\ep)Q_6 \,,
\end{eqnarray}
whereas for $\Gamma=v^\alpha$ we find $O_1^{\rm kin}=Q_5$, 
$O_T^{\rm kin}=Q_8$, $O_T^{\rm mag}=Q_{10}$, and
\begin{eqnarray}\label{rel2}
   O_1^{\rm mag} &=& -(1-\ep)(3-2\ep)Q_5 \,, \nonumber\\
   O_2^{\rm mag} &=& -(1-\ep)Q_5 \,.
\end{eqnarray}
Hence only for the case of insertions of the magnetic operator there 
appear nontrivial basis transformations. 

We thus consider the basis transformation from the six operators 
consisting of two sets of 
$\{O_1^{\rm mag},O_2^{\rm mag},O_T^{\rm mag}\}$ evaluated with 
$\Gamma=\gamma^\alpha$ and $\Gamma=v^\alpha$ to the new basis spanned
by the operators $\{Q_4,Q_5,Q_6,Q_9,Q_{10}\}$. The corresponding 
transformation matrices $\bm{R}(\ep)$ and $\bm{L}(\ep)$ can easily be
obtained from (\ref{rel1}) and (\ref{rel2}). It is then straightforward 
to compute the matrix $\bm{Z_C}$ corresponding to the operator mixing 
of $Q_{7\dots 10}$ into $Q_{4\dots 6}$, and then using the first 
relation in (\ref{magic}) to calculate the two-loop anomalous dimension 
matrix $\bm{\gamma_C}$. We find that
\begin{equation}\label{gammaC}
   \mbox{\boldmath$\gamma_C$\unboldmath}
   = \left( \begin{array}{ccc}
   \gamma_1^{\rm kin} & 0 & 0 \\
   0 & \gamma_1^{\rm kin} & 0 \\
   \gamma_1^{\rm mag}~ & \gamma_2^{\rm mag} & ~\gamma_3^{\rm mag} \\
   0 & \gamma_1^{\rm mag} + \gamma_2^{\rm mag} + \gamma_3^{\rm mag} & 0
   \end{array} \right) ,
\end{equation}
where
\begin{eqnarray}\label{results}
   \gamma_1^{\rm kin}
   &=& \frac{C_F\alpha_s}{4\pi} \Bigg\{ -8 + \frac{\alpha_s}{4\pi} 
    \Bigg[ C_F \left( - \frac{32}{3} - \frac{32\pi^2}{9} \right) 
    + C_A \left( - \frac{608}{9} + \frac{16\pi^2}{3} \right) 
    + \frac{160}{9}\,T_F\,n_f \Bigg] \Bigg\} \,, \nonumber\\
   \gamma_1^{\rm mag}
   &=& \frac{C_F\alpha_s}{4\pi} \Bigg\{ -2 + \frac{\alpha_s}{4\pi} 
    \Bigg[ C_F \left( - \frac{10}{3} - \frac{40\pi^2}{9} \right) 
    + C_A \left( \frac{46}{9} + \frac{10\pi^2}{9} \right) 
    - \frac{44}{9}\,T_F\,n_f \Bigg] \Bigg\} \,, \nonumber\\
   \gamma_2^{\rm mag}
   &=& \frac{C_F\alpha_s}{4\pi} \Bigg\{ \hspace{0.3cm} 12 + 
    \frac{\alpha_s}{4\pi} 
    \Bigg[ C_F \left( \frac{38}{3} + \frac{176\pi^2}{9} \right) 
    + C_A \left( \frac{236}{3} - \frac{44\pi^2}{9} \right) 
    - \frac{56}{3}\,T_F\,n_f \Bigg] \Bigg\} \,, \nonumber\\
   \gamma_3^{\rm mag}
   &=& \frac{C_F\alpha_s}{4\pi} \Bigg\{ -2 + \frac{\alpha_s}{4\pi} 
    \Bigg[ C_F \left( \frac{4}{3} - \frac{8\pi^2}{9} \right) 
    + C_A \left( - \frac{206}{9} + \frac{2\pi^2}{9} \right) 
    + \frac{52}{9}\,T_F\,n_f \Bigg] \Bigg\} \,.
\end{eqnarray}
This completes the calculation of the anomalous dimension matrix 
(\ref{gammamatr}) at two-loop order.

\section{Wilson coefficients for the vector current}
\label{sec:analysis}

The Wilson coefficients $B_j(\mu)$ in the heavy-quark expansion of the 
vector current in (\ref{vector}) obey the renormalization-group equation
\begin{equation}\label{RGE}
   \left( \frac{\mbox{d}}{\mbox{d}\ln\mu} - \bm{\gamma}^T \right)
   \vec B(\mu) = 0 \,,
\end{equation}
where the vector $\vec B(\mu)$ contains the ten coefficients $B_j(\mu)$.
The anomalous dimension matrix $\bm{\gamma}$ has been given in 
(\ref{gammamatr}), (\ref{entries}), (\ref{gammaA}), and (\ref{gammaC}).
Besides the entries of the submatrix $\bm{\gamma_C}$ computed in this 
paper and shown in (\ref{results}), we need the two-loop anomalous
dimension
\begin{equation}
   \gamma^{\rm mag}
   = \frac{C_A\alpha_s}{4\pi} \Bigg\{ 2 + \frac{\alpha_s}{4\pi} 
   \left[ \frac{68}{9}\,C_A - \frac{52}{9}\,T_F\,n_f \right] \Bigg\}
\end{equation}
of the chromo-magnetic operator calculated in \cite{ABN}, as well as
the two-loop anomalous dimensions of local dimension-4 currents,
\begin{eqnarray}
   \gamma_1
   &=& \frac{C_F\alpha_s}{4\pi} \Bigg\{ -3 + \frac{\alpha_s}{4\pi} 
    \Bigg[ C_F \left( \frac{5}{2} - \frac{8\pi^2}{3} \right) 
    + C_A \left( - \frac{49}{6} + \frac{2\pi^2}{3} \right) 
    + \frac{10}{3}\,T_F\,n_f \Bigg] \Bigg\} \,, \nonumber\\
   \gamma_2
   &=& \frac{C_F\alpha_s}{4\pi} \Bigg\{ \hspace{0.53cm} 3 +
    \frac{\alpha_s}{4\pi} 
    \Bigg[ C_F \left( -5 + \frac{4\pi^2}{3} \right) 
    + C_A \left( \frac{41}{3} - \frac{\pi^2}{3} \right) 
    - \frac{10}{3}\,T_F\,n_f \Bigg] \Bigg\} \,, \nonumber\\
   \gamma_3
   &=& \frac{C_F\alpha_s}{4\pi} \Bigg\{ -2 + \frac{\alpha_s}{4\pi} 
    \Bigg[ C_F \left( \frac{13}{3} - \frac{8\pi^2}{9} \right) 
    + C_A \left( - \frac{104}{9} + \frac{2\pi^2}{9} \right) 
    + \frac{28}{9}\,T_F\,n_f \Bigg] \Bigg\} \,, 
\end{eqnarray}
which were computed in \cite{Gabr}. The coefficient $\gamma_1$ coincides
with the universal hybrid anomalous dimension of the leading-order 
heavy--light currents first obtained in \cite{JiMu,BrGr}. In addition to 
the anomalous dimensions, we need the one-loop matching conditions for 
the Wilson coefficients at the scale $\mu=m_Q$. In the
$\overline{\mbox{MS}}$ subtraction scheme, they are \cite{MN94}
\begin{eqnarray}
   C_1(m_Q) &=& B_1(m_Q) = 1 - \frac{C_F\alpha_s}{\pi} \,,
    \nonumber\\
   C_2(m_Q) &=& B_2(m_Q) = \frac{1}{2}\,B_3(m_Q)
    = \frac{C_F\alpha_s}{2\pi} \,, \nonumber\\
   - B_4(m_Q) &=& \frac{1}{3}\,B_5(m_Q) = B_6(m_Q)
    = \frac{C_F\alpha_s}{\pi} \,, \nonumber\\
   C_{\rm mag}(m_Q) &=& 1 + (C_A+C_F)\,\frac{\alpha_s}{2\pi} \,.
\end{eqnarray}

The solution of the renormalization-group equation at next-to-leading
order is standard and described in detail in \cite{MN94}. A subtlety 
complicating the solution is that the one-loop anomalous dimension 
matrix cannot be diagonalized. We circumvent this problem by adding a
small diagonal contribution $\eta\,\bm{1}_{3\times 3}$ to the matrix 
$\bm{\gamma_B}$ in (\ref{gammamatr}), taking the limit $\eta\to 0$ at 
the end of the calculation.

We quote results for $N=3$ colors and $n_f=4$ light quark flavors. We
first reproduce the known NLO results
\begin{eqnarray}
   C_1(\mu) &=& B_1(\mu) = x^{6/25} \left[ 1
    - 6.243\,\frac{\alpha_s(m_Q)}{4\pi}
    + (0.910 - 2\Delta_{\rm RS})\,\frac{\alpha_s(\mu)}{4\pi} \right]
    \,, \nonumber\\
   C_2(\mu) &=& B_2(\mu) = \frac{1}{2}\,B_3(\mu) 
    = \frac{8}{3}\,x^{6/25}\,\frac{\alpha_s(m_Q)}{4\pi} \,,
    \nonumber\\
   C_{\rm mag}(\mu) &=& x^{-9/25} \left[ 1
    + 8.449\,\frac{\alpha_s(m_Q)}{4\pi}
    + (0.218 + 3\Delta_{\rm RS})\,\frac{\alpha_s(\mu)}{4\pi} \right] \,,
\end{eqnarray}
where $x=\alpha_s(\mu)/\alpha_s(m_Q)$. According to (\ref{B7to10}), 
products of these coefficients determine the coefficients 
$B_{7\dots 10}$. In addition, we obtain the new expressions
\begin{eqnarray}\label{Bi}
   B_4(\mu) &=& x^{6/25} \left[ \frac{34}{27}
    - 9.127\,\frac{\alpha_s(m_Q)}{4\pi}
    + \left( 2.820 - \frac{212}{27}\,\Delta_{\rm RS} \right)
    \frac{\alpha_s(\mu)}{4\pi} \right] \nonumber\\
   &+& x^{6/25} \ln x \left[ \frac{16}{25} 
    - 3.996\,\frac{\alpha_s(m_Q)}{4\pi}
    + \left( 0.582 - \frac{32}{25}\,\Delta_{\rm RS} \right)
    \frac{\alpha_s(\mu)}{4\pi} \right] \nonumber\\
   &+& x^{-3/25} \left[ - \frac{4}{27} 
    - 0.327\,\frac{\alpha_s(m_Q)}{4\pi}
    + \left( 2.013 - \frac{4}{27}\,\Delta_{\rm RS} \right)
    \frac{\alpha_s(\mu)}{4\pi} \right] \nonumber\\
   &-& \frac{10}{9} + 0.280\,\frac{\alpha_s(m_Q)}{4\pi}
    - 0.993\,\frac{\alpha_s(\mu)}{4\pi} \,, \nonumber\\
   B_5(\mu) &=& x^{6/25} \left[ - \frac{28}{27}
    + (32.097 + 1.707\ln x)\,\frac{\alpha_s(m_Q)}{4\pi}
    - \left( 0.944 - \frac{56}{27}\,\Delta_{\rm RS} \right)
    \frac{\alpha_s(\mu)}{4\pi} \right] \nonumber\\
   &+& x^{-3/25} \left[ \frac{88}{27} 
    + 11.930\,\frac{\alpha_s(m_Q)}{4\pi}
    - \left( 30.101 - \frac{88}{27}\,\Delta_{\rm RS} \right)
    \frac{\alpha_s(\mu)}{4\pi} \right] \nonumber\\
   &-& \frac{20}{9} + 0.560\,\frac{\alpha_s(m_Q)}{4\pi}
    + 2.458\,\frac{\alpha_s(\mu)}{4\pi} \,, \nonumber\\
   B_6(\mu) &=& x^{6/25} \left[ - 2 
    + 7.153\,\frac{\alpha_s(m_Q)}{4\pi}
    - (1.820 - 4\Delta_{\rm RS})\,\frac{\alpha_s(\mu)}{4\pi} \right]
    \nonumber\\
   &+& x^{-3/25} \left[ - \frac{4}{3} 
    - 2.941\,\frac{\alpha_s(m_Q)}{4\pi}
    + \left( 5.246 - \frac{4}{3}\,\Delta_{\rm RS} \right)
    \frac{\alpha_s(\mu)}{4\pi} \right] \nonumber\\
   &+& \frac{10}{3} - 0.841\,\frac{\alpha_s(m_Q)}{4\pi}
    - 1.464\,\frac{\alpha_s(\mu)}{4\pi} \,.
\end{eqnarray}
All terms proportional to the coupling $\alpha_s(m_Q)$ in these 
expressions, as well as the leading-logarithmic rescaling factors, are 
independent of the renormalization scheme. The terms proportional to
$\alpha_s(\mu)$ are scheme dependent, however. Here we consider a class 
of schemes parameterized by the quantity
\begin{equation}
   \Delta_{\rm RS}
   = \ln\frac{\mu_{\overline{\rm MS}}^2}{\mu_{\rm RS}^2} \,.
\end{equation}
These include all ``minimal-subtraction-like'' schemes, which are 
related to each other by a change in the renormalization scale. For 
instance, we have $\Delta_{\rm MS}=-\gamma_E+\ln 4\pi$ and 
$\Delta_{\overline{\rm MS}}=0$ by definition, i.e., the above results
evaluated with $\Delta_{\rm RS}=0$ refer to the $\overline{\mbox{MS}}$ 
scheme. It follows from (\ref{RGE}) that at NLO the scheme-dependent 
terms in the Wilson coefficients are given by
\begin{equation}
   \Delta\vec B(\mu) = \frac{\Delta_{\rm RS}}{2}\,\bm{\gamma}^T\,
   \vec B(\mu) \,.  
\end{equation}
In a complete NLO calculation, these terms combine with 
scheme-dependent terms in the hadronic matrix elements of the HQET 
operators to give a renormalization-group invariant answer.

\begin{figure}[t]
\epsfxsize=16cm
\centerline{\epsffile{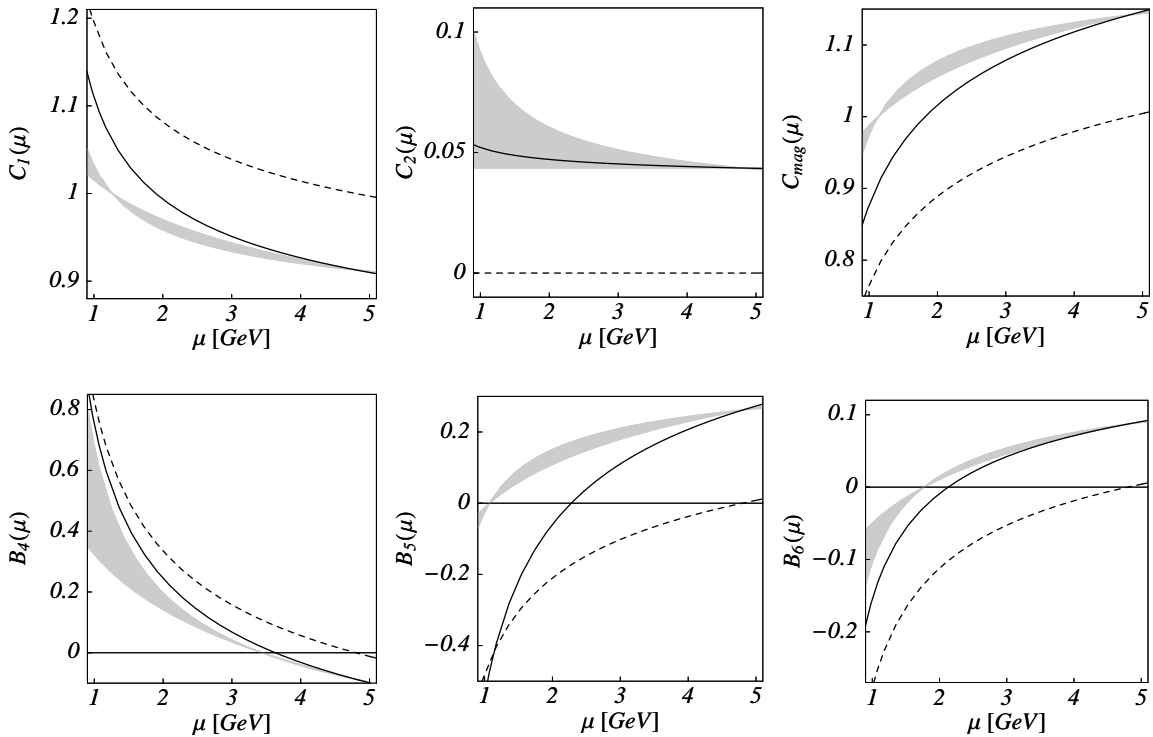}}
\centerline{\parbox{14.5cm}{\caption{\label{fig:2}
Different perturbative approximations for the Wilson coefficients in
the $\overline{\rm MS}$ scheme: next-to-leading order (solid), 
leading-logarithmic approximation (dashed), and one-loop order (band).}}}
\end{figure}

It is evident from the magnitude of some of the coefficients in 
(\ref{Bi}) that the next-to-leading corrections are, in some cases, 
quite large. To illustrate this point, we compare our NLO results with 
the naive one-loop expressions for the Wilson coefficients calculated 
in \cite{MN94}, and with the results obtained in leading-logarithmic 
approximation \cite{FaGr}, where all terms proportional to 
$\alpha_s/4\pi$ are omitted. The different results for the Wilson 
coefficients are shown in Figure~\ref{fig:2} as a function of the 
renormalization scale $\mu$, setting $m_Q=4.8$\,GeV corresponding to the
$b$-quark mass. For the one-loop approximation we show a band obtained 
by varying the coupling between $\alpha_s(\mu)$ and $\alpha_s(m_Q)$. We
note two important observations: (i) The one-loop and 
leading-logarithmic approximations give rather different results, which 
sometimes do not even have the same sign. This shows that the one-loop
matching corrections at the scale $\mu=m_Q$ are important. (ii) For
low renormalization scales $\mu\sim 1$--2\,GeV, some of the 
coefficients, in particular $B_4(\mu)$ and $B_5(\mu)$, are numerically 
large despite the fact that they vanish at tree level. This shows that 
the effects of running are important, and thus a RG summation of the 
large logarithms $\alpha_s\ln(m_Q/\mu)$ becomes mandatory. The effects 
of matching and running can only be combined in a consistent, 
scheme-independent manner by going beyond the leading order. Our NLO 
results (shown by the solid lines) typically lie between the two 
approximations, except for $\mu\approx m_Q$, where they agree with the 
naive one-loop results.

\section{Summary and an application}

We have calculated at two-loop order the mixing of bilocal operators 
into local dimension-4 current operators containing a heavy and a light 
quark field. When combined with results obtained by previous authors,
this completes the calculation of the $10\times 10$ two-loop anomalous 
dimension matrix governing the renormalization-scale dependence of the 
operators appearing at order $1/m_Q$ in the heavy-quark expansion of 
the weak current $\bar u\gamma^\mu(1-\gamma_5)b$. We have presented 
expressions for the Wilson coefficients in this expansion at
next-to-leading order in renormalization-group improved perturbation 
theory. We find that the next-to-leading corrections are numerically 
large and must be included for a meaningful determination of these
coefficients. This makes our results relevant for phenomenological 
applications of the heavy-quark expansion.

The matrix elements of dimension-4 heavy--light current operators play 
an important role in heavy-flavor phenomenology. They appear, e.g., at 
order $1/m_Q$ in the heavy-quark expansion of meson decay 
constants \cite{subl}, and of the semileptonic form factors describing 
the exclusive $\bar B\to\pi\,l\,\nu$ and $\bar B\to\rho\,l\,\nu$ 
transitions \cite{Burd}. The primary motivation for our calculation, 
however, was the recent proposal for extracting the element $|V_{ub}|$ 
of the quark mixing matrix from the lepton invariant mass spectrum in 
inclusive $B\to X_u\,l\,\nu$ decays \cite{Zolt}. This spectrum can be 
obtained using a two-step operator product expansion (called ``hybrid 
expansion'') of a time-ordered product of two heavy--light currents 
\cite{me}. Our results will help to reduce the perturbative uncertainty 
in this expansion, which will ultimately lead to an improved accuracy 
in the determination of $|V_{ub}|$.

At NLO in the hybrid expansion, one needs the expression for the 
combination of Wilson coefficients $(3B_4+B_6)$ derived in the present 
work. We now present an exact result for this combination, evaluating 
the coefficients with $N=3$ colors, but for an arbitrary number of 
light quark flavors. We find
\begin{eqnarray}
   3 B_4(\mu) + B_6(\mu)
   &=& \frac{16}{9}\,x^{2/\beta_0} \left[ 1
    + \frac{\alpha_s(m_Q)}{4\pi}\,(k_1+k_2)
    - \frac{\alpha_s(\mu)}{4\pi} \left( k_1 - \frac{47}{6}
    + 11\Delta_{\rm RS} \right) \right] \nonumber\\
   &-& \frac{16}{9}\,x^{-1/\beta_0} \left[ 1
    + \frac{\alpha_s(m_Q)}{4\pi}\,k_3
    - \frac{\alpha_s(\mu)}{4\pi} \left( k_3-k_2 - \frac{34}{3}
    - \Delta_{\rm RS} \right) \right] \nonumber\\
   &+& \frac{16}{\beta_0}\,x^{2/\beta_0} \ln x \left[ 1
    + \frac{\alpha_s(m_Q)}{4\pi}\,k_4
    - \frac{\alpha_s(\mu)}{4\pi} \left( k_4 + \frac{16}{3} 
    + 2\Delta_{\rm RS} \right) \right] \nonumber\\
   &-& \frac{40}{9}\,\frac{\alpha_s(\mu)}{4\pi} \,,
\end{eqnarray}
where
\begin{eqnarray}\label{ki}
   k_1 &=& - \frac{53}{6} + \left( \frac{1208}{9} + \frac{314\pi^2}{27}
    \right) \frac{1}{\beta_0} - \frac{1177}{\beta_0^2} 
    \simeq 4.098 \,, \nonumber\\
   k_2 &=& - \left( \frac{311}{6} + \frac{28\pi^2}{9} \right)
    \frac{1}{\beta_0-3} \simeq -15.476 \,, \nonumber\\
   k_3 &=& \frac{49}{6} - \left( \frac{941}{18} + \frac{28\pi^2}{27}
    \right) \frac{1}{\beta_0} + \frac{107}{\beta_0^2} 
    \simeq 2.206 \,, \nonumber\\
   k_4 &=& - 7 + \left( \frac{380}{9} - \frac{28\pi^2}{27}
    \right) \frac{1}{\beta_0} - \frac{214}{\beta_0^2} 
    \simeq -6.243 \,,
\end{eqnarray}
and $\beta_0=11-\frac23 n_f$ is the first coefficient of the 
$\beta$-function. The numerical values in (\ref{ki}) refer to $n_f=4$.
In \cite{newwork}, we combine this result with an explicit calculation 
of the hadronic matrix elements in the hybrid expansion, finding that
the scheme dependence parameterized by $\Delta_{\rm RS}$ disappears from 
the final result for the semileptonic $B\to X_u\,l\,\nu$ decay rate.

\vspace{0.3cm}\noindent
{\it Acknowledgements:\/}
One of us (M.N.) would like to thank Gabriel Amor\'os for earlier
collaboration on a similar subject. This work was supported in part by 
the National Science Foundation. T.B.~is supported by the Swiss National
Science Foundation.


\begin{thebibliography}{99}
\parskip=0pt

\bibitem {review}
For a review, see: M. Neubert, Phys.\ Rep.\ {\bf 245}, 259 (1994).

\bibitem {FaGr}
A.F. Falk and B. Grinstein, Phys.\ Lett.\ B {\bf 247}, 406 (1990).

\bibitem {MN94}
M. Neubert, Phys.\ Rev.\ D {\bf 49}, 1542 (1994).

\bibitem {ABN}
G. Amor\'os, M. Beneke and M. Neubert, Phys.\ Lett.\ B {\bf 401}, 81
(1997).

\bibitem {LuMa}
M. Luke and A.V. Manohar, Phys.\ Lett.\ B {\bf 286}, 348 (1992).

\bibitem {JiMu}
X. Ji and M.J. Musolf, Phys.\ Lett.\ B {\bf 257}, 409 (1991).

\bibitem {BrGr}
D.J. Broadhurst and A.G. Grozin, Phys.\ Lett.\ B {\bf 267}, 105
(1991).

\bibitem{Gabr}
G. Amor\'os and M. Neubert, Phys.\ Lett.\ B {\bf 420}, 340 (1998).

\bibitem {Flor}
E.G. Floratos, D.A. Ross and C.T. Sachrajda, Nucl.\ Phys.\ B {\bf 129},
66 (1977).

\bibitem {Che}
K.G. Chetyrkin and F.V. Tkachov, Phys.\ Lett.\ B {\bf 114}, 340 (1982);\\
K.G.~Chetyrkin and V.A.~Smirnov, Phys.\ Lett.\ B {\bf 144}, 419 (1984).

\bibitem {Chet}
K.G. Chetyrkin and F.V. Tkachov, Nucl.\ Phys.\ B {\bf 192}, 159
(1981).

\bibitem {subl}
M. Neubert, Phys.\ Rev.\ D {\bf 46}, 1076 (1992).

\bibitem {Burd}
G. Burdman, Z. Ligeti, M. Neubert and Y. Nir, Phys.\ Rev.\ D {\bf 49},
2331 (1994).

\bibitem{Zolt}
C.W. Bauer, Z. Ligeti and M. Luke, Phys.\ Lett.\ B {\bf 479}, 395 
(2000).

\bibitem{me}
M. Neubert, J.\ High Ener.\ Phys.\ 0007, 022 (2000).

\bibitem{newwork}
T. Becher and M. Neubert, {\it Improved Determination of $|V_{ub}|$ from 
Inclusive Semileptonic $B$-Meson Decays}, Preprint CLNS~01/1737 
[{\tt hep-ph/0105217}].

\end{thebibliography}
\end{document}